\documentclass{aa}
\usepackage{ulem}
\usepackage{graphicx}
\usepackage{txfonts}
\usepackage{xcolor}
\usepackage{xspace}
\usepackage[colorlinks=true, citecolor=blue, linkcolor=blue]{hyperref}
\usepackage{perpage} 
\MakePerPage{footnote}
\usepackage{arydshln}
\usepackage[switch]{lineno} 

\newcommand{\courrier}[1]{{\fontfamily{lmtt}\selectfont{#1}}}
\newcommand{\ita}[1]{{\textit{#1}}}
\newcommand{\fermi}{\ita{Fermi}-LAT\xspace}
\newcommand{\xmm}{\ita{XMM-Newton}\xspace}
\newcommand{\chandra}{\ita{Chandra}\xspace}
\newcommand{\swift}{\ita{Swift}\xspace}

\newcommand{\asca}{\ita{ASCA}\xspace}
\newcommand{\suzaku}{\ita{Suzaku}\xspace}

\newcommand{\rspec}{$R_{\rm{spec}}$\xspace}

\newcommand{\dg}{$^{\circ}$\xspace}

\newcommand{\dfu}{erg cm$^{-2}$ s$^{-1}$\xspace}

\newcommand{\gui}[1]{{``{#1}''}}
\newcommand{\hii }{H\,{\sc ii}\xspace}
\newcommand{\hi }{H\,{\sc i}\xspace}

\titlerunning{Multiwavelength constraints on unidentified TeV sources}
\authorrunning{J. Devin et al.}

\begin{document}

\title{Multiwavelength constraints on the unidentified Galactic  \\ TeV sources HESS~J1427$-$608, HESS~J1458$-$608, \\ and new VHE $\gamma$-ray source candidates}

    \author{J. Devin\inst{1}
            \and M. Renaud\inst{2}
            \and M. Lemoine-Goumard\inst{1}
            \and G. Vasileiadis\inst{2}
          }

   \institute{Univ. Bordeaux, CNRS, CENBG, UMR 5797, F-33170 Gradignan, France \\
        \email{jdevin.phys@gmail.com, devin@cenbg.in2p3.fr}
         \and
            Laboratoire Univers et Particules de Montpellier, CNRS/IN2P3, Universit\'e de Montpellier, F-34095 Montpellier, France 
            }

   \date{Received 30 September 2020, Accepted 6 January 2021}

\abstract
  {}
   {Among the $\gamma$-ray sources discovered at high and very-high energies, a large fraction still lack a clear identification. In particular, the H.E.S.S. Galactic Plane Survey (HGPS) revealed 78 TeV sources among which 47 are not clearly associated with a known object. Multiwavelength data can help identify the origin of the very-high energy $\gamma$-ray emission, although some bright TeV sources have been detected without clear counterparts. We present a multiwavelength approach to constrain the origin of the emission from unidentified HGPS sources.}
   {We present a generic pipeline that explores a large database of multiwavelength archival data toward any region in the Galactic plane. Along with a visual inspection of the retrieved multiwavelength observations to search for faint and uncataloged counterparts, we derive a radio spectral index that helps disentangle thermal from nonthermal emission and a mean magnetic field through X-ray and TeV data in case of a leptonic scenario. We also search for a spectral connection between the GeV and the TeV regimes with the \fermi cataloged sources that may be associated with the unidentified HGPS source. We complete the association procedure with catalogs of known objects (supernova remnants, pulsar wind nebulae, \hii regions, etc.) and with the source catalogs from instruments whose data are retrieved.}
   {The method is applied on two unidentified sources, namely HESS~J1427$-$608 and HESS~J1458$-$608, for which the multiwavelength constraints favor the pulsar wind nebula (PWN) scenario. We model their broadband nonthermal spectra in a leptonic scenario with a magnetic field $B \lesssim 10$ $\mu$G, which is consistent with that obtained from ancient PWNe. We place both sources within the context of the TeV PWN population to estimate the spin-down power and the characteristic age of the putative pulsar. We also shed light on two possibly significant $\gamma$-ray excesses in the HGPS: the first is located in the south of the unidentified source HESS~J1632$-$478 and the second is spatially coincident with the synchrotron-emitting supernova remnant G28.6$-$0.1. The multiwavelength counterparts found toward both $\gamma$-ray excesses make these promising candidates for being new very-high energy $\gamma$-ray sources.}
   {}


   \maketitle

\section{Introduction}

Supernova remnants (SNRs) and pulsar wind nebulae (PWNe) are considered the best candidates to accelerate the bulk of Galactic cosmic rays (CRs) at least up to the knee of the CR spectrum ($\sim 3 \times 10^{15}$ eV). Although no evidence for particle acceleration up to PeV energies has been found so far, efficient particle acceleration in SNRs and PWNe has been observed and can be explained by the widely adopted diffusive shock acceleration mechanism \citep[DSA;][]{Bell:1978}. Accelerated electrons and positrons radiate synchrotron emission and also $\gamma$ rays through Bremsstrahlung and inverse Compton (IC) scattering on photon fields while accelerated protons and heavier nuclei colliding with ambient matter emit $\gamma$ rays through neutral pion decay. Thus, $\gamma$-ray studies probe the population of both electrons and protons and allow us to disentangle different origins of the emission. However, $\gamma$-ray instruments usually have lower angular resolution than most radio and X-ray instruments, and their observations often suffer from source confusion. Thus, probing the accelerated leptons through the expected synchrotron emission mainly in the radio and X-ray bands helps constrain the nature of  $\gamma$-ray sources. 

The identification of $\gamma$-ray emitting sources in the high-energy (HE;  0.1 $<$ $E$ $<$ 100 GeV) and very-high energy (VHE; 0.1 TeV $<$ $E$ $<$ 100 TeV) domains often relies on spatial correlation with identified objects at other wavelengths. While catalogs of known $\gamma$-ray emitters in the Galaxy such as SNRs, PWNe, and pulsars can provide hints of the origin of the emission, multiwavelength data exploration is often necessary to pinpoint the nature of the HE/VHE sources \citep{VERITAS_TeV2032:2014, MAGIC_1857:2014, MAGIC_1841:2020}. For instance, the TeV shell HESS~J1534$-$571 was found to be spatially coincident with the SNR candidate G323.7$-$1.0 (discovered through a close inspection of archival radio images) and was thus identified as the VHE counterpart of this SNR \citep{TeVshell:2018}. One of the brightest TeV sources, HESS~J1825$-$137, was also identified as a PWN owing to its association with the X-ray PWN G18.0$-$0.7 and its $\gamma$-ray energy-dependent morphology shrinking to the energetic pulsar PSR~J1826$-$1334 at the highest energies \citep{1825:2005}. Finally, the source HESS~J1356$-$645 was classified as an evolved PWN as a result of the exploitation of multiwavelength data that revealed a nonthermal, center-filled radio and X-ray emission surrounding the energetic pulsar PSR~J1357$-$6429 and spatially coincident with the H.E.S.S. source \citep{HESSJ1356_2011}. 

Several databases have been developed to gather multiwavelength observations that can be easily retrieved. The \ita{Centre d'Analyse des Donn\'{e}es \'{E}tendues}\footnote{\url{http://cade.irap.omp.eu/dokuwiki/doku.php?id=start}} and the \ita{Space Science Data Center}\footnote{\url{http://www.ssdc.asi.it/mma.html}} provide interactive databases, such as the \ita{Skyview}\footnote{\url{https://skyview.gsfc.nasa.gov/current/cgi/titlepage.pl}} tool, which enables the production of images at multiple wavelengths for any position in the sky. The interactive tool \ita{gamma sky}\footnote{\url{http://gamma-sky.net/map}} permits us to superimpose catalogs such as those of the \fermi and known SNRs. 

However, using multiwavelength data to identify $\gamma$-ray sources is not always straightforward. The best example is the nearby and evolved PWN Vela X, which exhibits a changing morphology at different wavelengths. The brightest part of the VHE emission is spatially coincident with the X-ray PWN, but its largest extent is consistent with that of the surrounding extended radio nebula of 2\dg $\times$ 3\dg \citep{VelaX_HESS2012}. Emission in the north of the pulsar is also seen with the \fermi \citep{Grondin:2013}, illustrating how the interpretation of multiwavelength data can be difficult and puzzling. Moreover, multiwavelength associations relying on spatial correlations is challenging if the $\gamma$-ray emission is produced by an ancient PWN or a molecular cloud (MC) illuminated by CRs that have escaped from a nearby SNR. In the former case, the pulsar can be largely offset from the TeV emission produced by old electrons \citep{deJager:2009} and, in the latter case, the MC can also be distanced from the SNR from which CRs have escaped \citep{Gabici:2009}. Thus, there may be no spatial correlation between the driven-phenomenon of the emission (here pulsar or SNR) and the TeV emission itself, making the association with multiwavelength data difficult. Finally, nonthermal radio and X-ray emission from ancient PWNe and MCs illuminated by CRs is expected to be faint, challenging the identification process. 
 
 With 2700 hours of observations, the H.E.S.S. experiment, through a Galactic Plane Survey \citep[HGPS;][]{HGPS:2018} covering the inner part of the Galaxy ($l = 250$\dg to $65$\dg and $\mid$b$\mid < 3.5$\dg), has reported 78 TeV sources, among which 12 PWNe, 16 SNRs (or composite SNRs), 3 binaries, and 47 sources that are not firmly identified. With a point spread function (PSF) of $\sim$ 0.08\dg (68\% containment radius) and a point-source sensitivity of $\lesssim$ 1.5\% of the Crab Nebula VHE flux, this Galactic scan was performed from 0.2 to 100 TeV with an unprecedented spatial coverage. Besides positional evidence with cataloged objects, the firm identification of the VHE sources was based on correlated multiwavelength variability, matching multiwavelength morphology and energy-dependent $\gamma$-ray morphology. The association procedure made use of the 3FGL and 2FHL \fermi catalogs \citep{3FGL, 2FHL}, the \gui{SNRcat} \citep{Ferrand:SNRcat}, the ATNF pulsar catalog \citep[][version 1.54]{ATNF:2005}, and 20 external analyses involving data at other wavelengths. When a firm identification was not possible, the H.E.S.S. sources were associated with all objects whose cataloged position is at an angular distance smaller than the H.E.S.S. source spectral extraction radius (noted \rspec). For a total of 47 unidentified sources, the HGPS association procedure reported 11 sources that are not associated with any cataloged sources and 36 sources for which the origin of the emission is unclear, mainly owing to several possible scenarios and source confusion.
 
Given the large number of TeV sources without firm identification, we developed a generic pipeline that retrieves archival multiwavelength data toward any region of the Galactic plane to constrain the origin of the $\gamma$-ray emission. Although extragalactic sources can be located near the Galactic plane \citep[as exemplified by the blazar HESS~J1943+213, ][and references therein]{HGPS:2018}, we are primarily interested in Galactic sources that are either extended at TeV energies or associated with an extended component seen at other wavelengths. Thus, in this study we also discarded high-mass binary systems, some of which are known to be HE and VHE emitters such as PSR~B1259$-$63 \citep[e.g.,][]{Chernyakova:2020}. We first describe our multiwavelength approach in Section~\ref{sec:MWL}. In Section~\ref{sec:1427_1458}, we apply this method on two unidentified sources, namely HESS~J1427$-$608 and HESS~J1458$-$608, and place them within the context of the PWN population seen at TeV energies. Finally, Section~\ref{sec:targets_CTA} sheds light on possibly significant (but not cataloged) TeV excesses in two confused regions: in the south of HESS~J1632$-$478 and toward the SNR G28.6$-$0.1, which both have multiwavelength counterparts, strengthening the interest in carrying out dedicated $\gamma$-ray analyses to confirm these as new VHE sources.

\section{Multiwavelength approach}\label{sec:MWL}

We present in this section a generic pipeline that retrieves archival radio, X-ray, infrared, and GeV data to search for uncataloged, presumably faint counterparts of the TeV unidentified sources. The association procedure is made with the instrument source catalogs whose data are retrieved and with catalogs of known objects that are more numerous than those used for the HGPS association procedure. After automatically downloading all the archival data and performing the association procedure, we took prior attention to the visual inspection of all the archival multiwavelength images to search for faint excesses within the TeV source extent and to define the regions of interest for flux extraction. This is the only step that needs manual intervention. After this, the pipeline automatically derives a radio spectral index that allows us to disentangle thermal from nonthermal emission and a mean magnetic field using X-ray and TeV data, assuming that the $\gamma$-ray emission is produced by IC scattering off the cosmic microwave background (CMB) photon field. Finally, the pipeline automatically plots the spectra of the \fermi cataloged sources found within the TeV source extent together with that of the H.E.S.S. source to search for a spectral connection between the HE and VHE regimes.

\subsection{Archival data retrieval}

Given a position in the sky and a search radius, we developed a code that automatically extracts radio continuum data from 13 single-dish telescopes or interferometers that survey the northern and the southern sky. This includes the second epoch of the Molongo Galactic Plane Survey \citep[MGPS-2, 843 MHz;][]{MGPS2_Murphy:2007}, the VLA Galactic Plane Survey \citep[VGPS, 1.4 GHz;][]{VGPS_Stil:2006}, the Southern Galactic Plane Survey \citep[SGPS, 1.4 GHz, a combination of the Australia Telescope Compact Array and the Parkes Radio Telescope;][]{SGPS_McClure:2005}, the TIFR GMRT Sky Survey \citep[TGSS, 150 MHz;][]{TGSS_Intema:2017}, the NRAO-VLA Sky Survey \citep[NVSS, 1.4 GHz;][]{NVSS_Condon:1998}, the Canadian Galactic Plane Survey \citep[CGPS, 1.4 GHz;][]{CGPS_Taylor:2003}, the Green Bank Telescope \citep[8.35 at 14.35 GHz and the 87GB survey at 4.85 GHz;][]{GBT_GPA_Langston:2000,87GB_Gregory:1991}, and the HI/OH/Recombination line survey of the inner Milky Way \citep[THOR, 1, 1.3, 1.4, 1.6, and 1.8 GHz;][]{THOR_Beuther:2016}. We also used data from the Parkes 64 m radio telescope \citep[2.4 GHz;][]{Parkes_Duncan:1995} and those from the Parkes-MIT-NRAO survey\footnote{This survey was made using the Parkes 64 m radio telescope with the NRAO multibeam receiver at a frequency of 4.85 GHz.} \citep[PMN, 4.85 GHz;][]{PMN_Griffith:1993}. Archival data from the HI Parkes All-Sky Survey (HIPASS) and the HI Zone of Avoidance (HIZOA) survey were reprocessed into a new continuum map that we retrieved \citep[CHIPASS, 1.4 GHz;][]{CHIPASS_Calabretta:2014}. We also retrieved data from the Multi-Array Galactic Plane Imaging Survey \citep[MAGPIS, 0.325, 1.5 and 5 GHz;][]{MAGPIS_Helfand:2006}. The technical details of the 13 radio instruments and data retrieval are given in Appendix~\ref{sec:Appendix_A} (and references therein). In addition, data from the Planck satellite are also used \citep[30, 44, 70, 100, and 143 GHz, ][]{Planck:2016}.

In the X-ray range, we retrieved data from \chandra/ACIS (0.5--7 keV), \xmm (0.2--12 keV), ASCA (0.4$-$10, 0.7$-$10 keV), {\it Integral}/IBIS-ISGRI (17$-$60 keV), \swift/XRT (0.2--10 keV), \swift/BAT (15--150 keV), NuSTAR (5--80 keV), \suzaku (0.2--12 keV), and ROSAT/PSPC (0.1--2.4, 0.5--2.0, 0.1--0.4, 0.4--0.9 and 0.9--2.4 keV). We automatically extracted all the observations whose center is comprised within an angular distance from the source of interest. Except for large field-of-view instruments (such as ROSAT/PSPC, {\it Integral}/IBIS-ISGRI and \swift/BAT), we built a mosaic of these images in case of multiple observations. When possible, we created background-subtracted and exposure-corrected images. Appendix~\ref{sec:Appendix_A} gives the technical description of the X-ray instruments and data retrieval.

We completed our data set with infrared data from Spitzer/GLIMPSE (3.6, 4.5, 5.8, 8, 21, 24, 870, and 1100 $\mu$m) and \fermi data (see Appendix~\ref{sec:Appendix_A} for more details).

For the association procedure, we used catalogs of objects reporting pulsars, PWNe, SNRs, \hii regions, etc. This includes the ATNF pulsar catalog \citep[][version 1.58]{ATNF:2005}, the \gui{SNRcat} \citep{Ferrand:SNRcat}, the Galactic SNR catalog \citep{Green:2017}, and the catalog of 76 Galactic SNR candidates revealed with THOR data \citep{Anderson:2017}. We used \fermi pulsar, point-source, and extended-source catalogs  (2PC, 3FGL, 2FHL, 3FHL, FGES, and 4FGL catalogs\footnote{References: \cite{2PC:2013, 3FGL, 2FHL, 3FHL:2017, FGES:2017, 4FGL}}) as well as the HAWC catalog \citep[2HWC;][]{2HWC:2017}. We note that energetic pulsars are considered indirect associations since we do not expect them to produce TeV emission but rather to generate a PWN that could be seen at TeV energies. We also used the catalogs of \hii regions obtained with WISE data \citep{Anderson_WISE:2014}, of MCs in the Milky Way \citep{Rice:2016}, of Galactic O stars \citep{GOSC:2013}, and information reported in the TeV source catalog\footnote{\gui{TeVcat}: \url{http://tevcat2.uchicago.edu}}. The details of the catalogs are given in Appendix~\ref{sec:Appendix_A}. To reduce potential missing associations, we automatically requested the Simbad\footnote{\url{http://simbad.u-strasbg.fr/simbad/}} database, which provides information on the listed astronomical objects.

\subsection{Nonthermal radio emission}

When extended radio emission is found toward the unidentified TeV source, we calculated a radio spectral index ${\alpha}$ (defined as $S_{\nu} \propto \nu^{\alpha}$, where $S_{\nu}$ and $\nu$ are the flux density and frequency, respectively) that helps identify thermal from nonthermal (i.e., synchrotron) emission. While $\alpha \gtrsim$ 0 in thermal sources such as \hii regions, its average value amounts to $\sim$ $-$0.4/$-$0.5 in SNRs \citep{Green:2017} and $\sim$ $-$0.3/0 in PWNe \citep{deJager:2009}. Although radio emission from PWNe and \hii regions can exhibit a similar spectral index, they differ in shape with a shell-like and a Gaussian-like morphology for \hii regions and PWNe, respectively. In each archival image, we masked the cataloged radio sources to estimate the noise and we calculated the flux in a given region of interest (called $R_{\rm{ON}}$) by summing up the flux in the pixels and correcting from the instrument beam. The flux extraction region was chosen after a thorough visual inspection of the data. The background-corrected fluxes, calculated at different frequencies, were fit with a power law to derive the radio spectral index. Upper limits on flux were derived at the 3$\sigma$ confidence level. We followed the same method as for the PWN HESS~J1356$-$645 \citep{HESSJ1356_2011}, which is now implemented in a generic way using all the retrieved radio observations. Detailed explanations on the radio spectral index derivation, with an illustration on the PWN HESS~J1356$-$645, are given in Appendix~\ref{sec:Appendix_B}.

\subsection{Mean magnetic field}

Nonthermal X-ray emission probes the presence of the highest-energy electrons that also radiate TeV $\gamma$ rays through IC scattering on photon fields. Although ROSAT/PSPC detects soft X-ray photons (up to 2.4 keV), which can be heavily absorbed by the interstellar gas, we took advantage of its large field of view ($\sim$ $6^{\circ} \times 6^{\circ}$) to derive constraints on the X-ray flux within any spectral extraction region defined in the HGPS. We masked the cataloged sources \citep[reported in the 1RXS and 2RXS catalogs;][]{1RXS:1999, 2RXS:2016} to estimate the background with the \ita{ring} and \ita{reflected background methods} \citep{Berge:2007}. The significance is calculated following \cite{Li_Ma:1983}, taking a correlation radius equal to the spectral extraction radius (\rspec) used in the HGPS. For a given X-ray spectral index $\Gamma_{\rm{X}}$ and a column density $N_{\rm{H}}$, the source count rate (or its 5$\sigma$ upper limit\footnote{In case there is no significant ($> 5 \sigma$) detection, we simulate ten X-ray sources at a given count rate and with the same morphology as that found in the HGPS. The significance of each of these simulated sources, once added to the ROSAT image, is calculated by applying the same methods as described in the text. After repeating this procedure for several source count rates, the 5$\sigma$ upper limit is derived when it leads to a mean significance of 5.}) in the 0.9$-$2.4 keV band is converted into a flux value (or upper limit) using the HEASARC tool WebPimms\footnote{\url{https://heasarc.gsfc.nasa.gov/Tools/w3pimms_help.html}}. Assuming a one-zone model and that the TeV emission is produced by IC scattering on the CMB in the Thomson regime ($\Gamma_{\rm{X}} = \Gamma_{\rm{TeV}} = \Gamma$), the X-ray and TeV flux measurements ($F_{\rm{sync}}$, $F_{\rm{IC}}$) can be used to constrain the mean magnetic field, which is expressed as
\begin{equation}
\frac{B}{10 \hspace{0.1cm} \mu G} \propto \bigg (\frac{F_{\rm{sync}}}{F_{\rm{IC}}} \times \frac{E_{\rm{2,IC, TeV}}^{2-\Gamma} - E_{\rm{1,IC, TeV}}^{2-\Gamma}}{E_{\rm{2,sync, keV}}^{2-\Gamma} - E_{\rm{1,sync, keV}}^{2-\Gamma}} \bigg )^{1/ \Gamma}
\label{eq:B_field}
,\end{equation}
where $\Gamma$ = ($p$ + 1)/2 is the photon spectral index and $p$ is the particle spectral index. More details on the mean magnetic field estimate are given in Appendix~\ref{sec:Appendix_C}, with an illustration on the SNR RX~J1713.7$-$3946. Although nonthermal X-ray filaments detected in several SNRs imply a magnetic field amplification up to hundreds of $\mu$G \citep{Parizot:2006}, the expected mean magnetic field value in case of linear DSA is roughly $B \sim 15-20$  $\mu$G for SNRs \citep{Renaud:2009} and $B \sim 3-8$ $\mu$G for evolved PWNe \citep{Torres:2014}.

\subsection{Gamma-ray spectra}

The GeV spectra show different features depending on the origin of the emission. $\gamma$-ray emission from pulsars is usually described by a power law with an exponential cutoff occurring below tens of GeV or by a logarithmic parabola. The second \fermi pulsar catalog \citep[2PC; ][]{2PC:2013} contains 117 pulsars for 2796 radio-detected pulsars. This difference can be explained by the pulsar spin-down power $\dot{E}$, which needs to be high enough to provide efficient particle acceleration up to $\gamma$-ray emitting energies. The emission from PWNe is usually characterized by a power law with a hard spectral shape ($\Gamma < 2$, where $\Gamma$ is the photon spectral index) originating from IC scattering on photon fields. This hard spectral shape limits the detection of PWNe with the \fermi and makes PWNe the most numerous VHE-emitting objects in our Galaxy. While a hard spectral shape is indicative of a leptonic origin of the emission, a roll-off in the photon spectrum below 100$-$200 MeV is the signature of accelerated protons colliding with ambient matter. In this case, the photon spectrum at $\gtrsim$ GeV energies is usually found to be soft ($\Gamma \gtrsim 2$) in SNRs interacting with MCs such as IC 443, W44, or W51C \citep[][and references therein]{Jogler:2016}. Given the crucial information brought by GeV spectra, we retrieved the \fermi spectrum of all cataloged sources located within the \rspec of the HGPS unidentified source to search for a spectral connection between GeV and TeV energies.

\section{Application on HESS~J1427$-$608 and HESS~J1458$-$608}\label{sec:1427_1458}

\subsection{HESS~J1427$-$608}\label{sec:HESSJ1427}

\begin{figure*}[th!]
    \centering
    \includegraphics[scale=0.43]{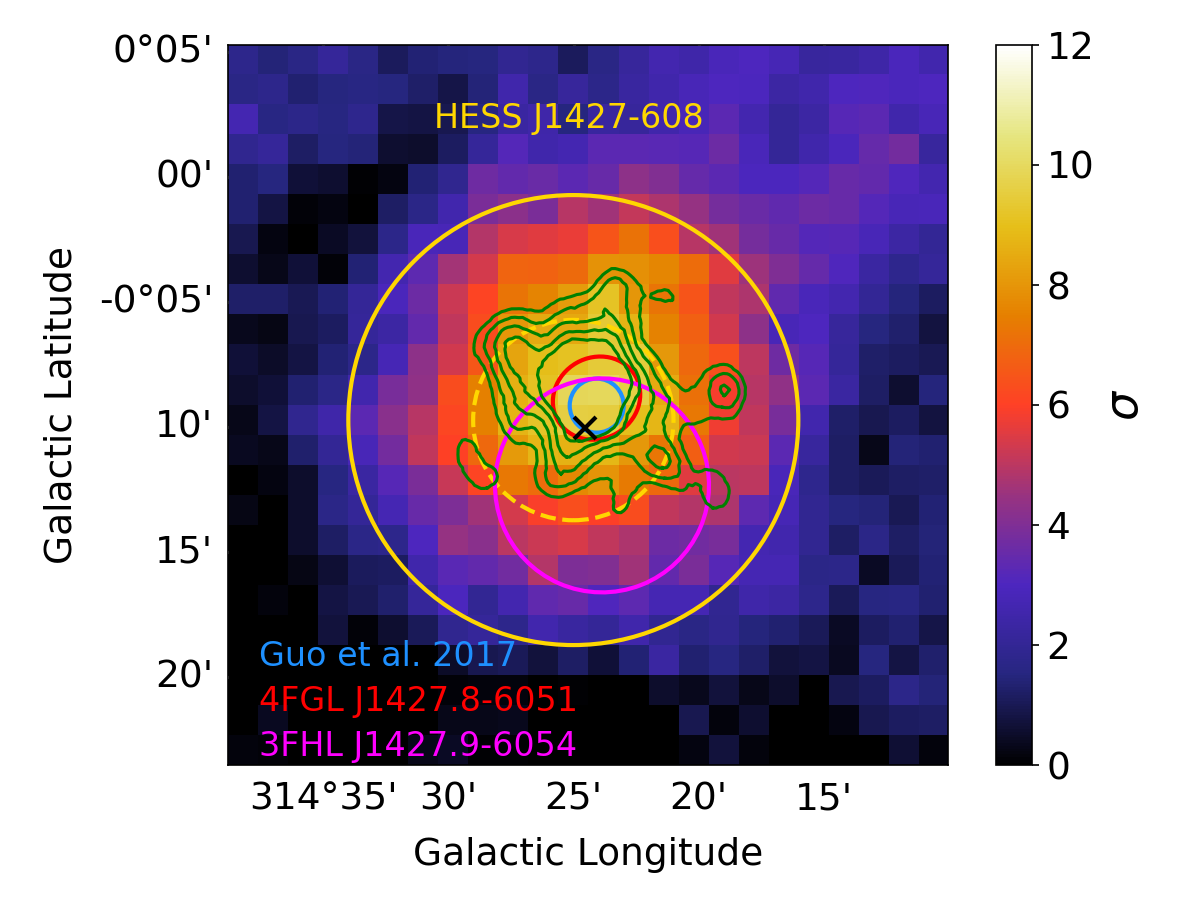} 
    \includegraphics[scale=0.43]{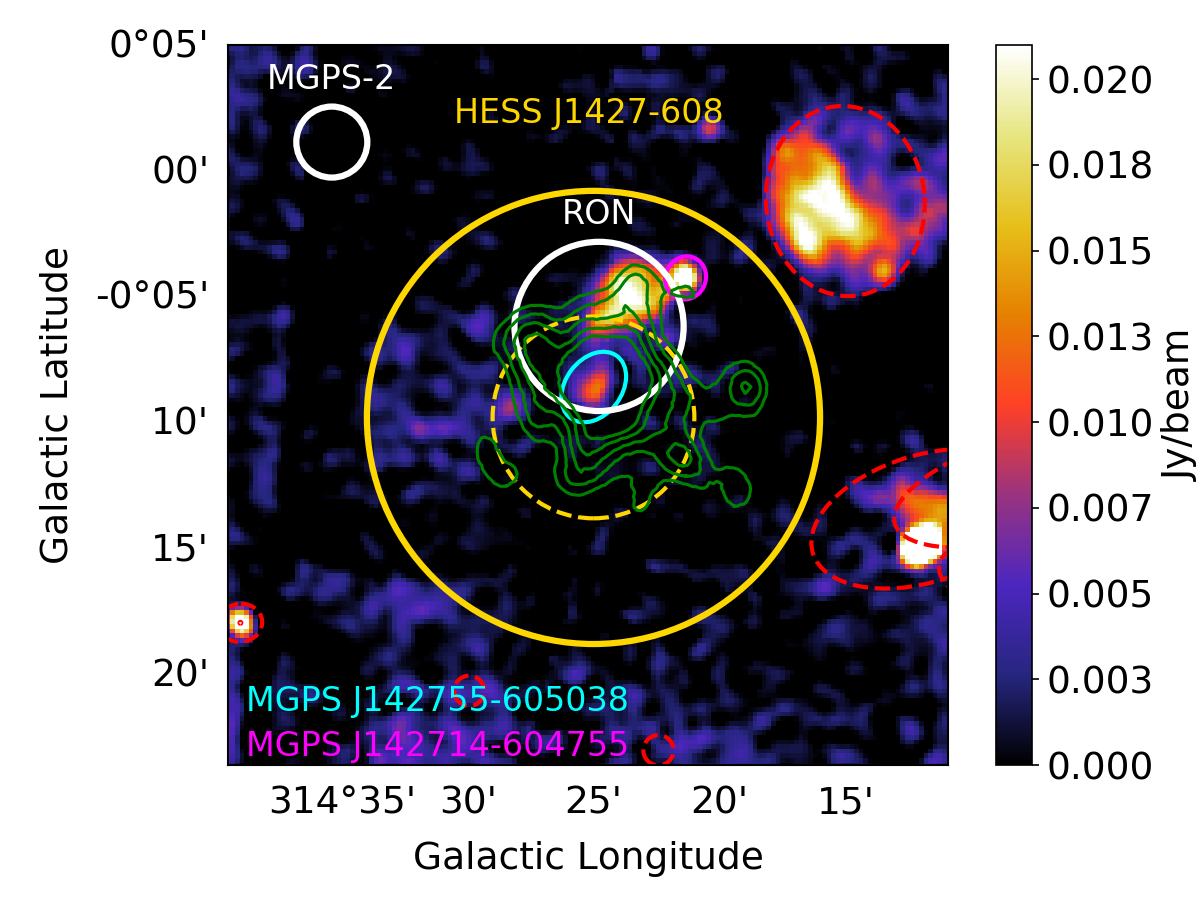} 
    \caption{(Left) HGPS significance map (obtained with a correlation radius of 0.1\dg) with the \fermi cataloged sources (within 95\% uncertainties) overlaid. The blue circle corresponds to the position uncertainty (1$\sigma$) of the \fermi source detected by \cite{Guo:2017} above 3 GeV. The H.E.S.S. upper limit on the TeV source extent and the spectral extraction radius \rspec are represented by the dashed and solid yellow circles, respectively. The green contours represent the emission seen by \suzaku between 2 keV and 8 keV \citep[Suzaku~J1427$-$6051,][]{Fujinaga:2013}. The black cross corresponds to the position of 4XMM~J142756.7$-$605214 (see Section~\ref{sec:1427_MWL}). (Right) MGPS-2 map (843 MHz) toward HESS~J1427$-$608. The red ellipses correspond to the MGPS-2 and PMN cataloged sources outside \rspec, while the cyan and magenta ellipses represent the two MGPS-2 sources inside \rspec. The FWHM of the MGPS-2 is given at the top left of the image. The radio flux extraction region is represented by a white circle (called $R_{\rm{ON}}$). The yellow circles and green contours are the same as those in the left panel.}
    \label{fig:HESSJ1427_maps}
\end{figure*}

\subsubsection{Source presentation}\label{sec:HESSJ1427_presentation}

HESS~J1427$-$608 is not spatially resolved in the HGPS and is therefore defined as a point source with a significance of $\sqrt{\rm{TS}}$ $\sim 10.5$ \citep{HGPS:2018}\footnote{The test statistic (TS) is the ratio of the logarithm of two likelihoods obtained with and without the source model. The TS follows a $\chi^{2}$ distribution with $n$ additional degrees of freedom.}. Using a two-dimensional symmetric Gaussian, \cite{Aharonian:2008} derived a size of $\sigma$ = 0.063\dg $\pm$ 0.01\dg while the HGPS reported $\sigma$ = 0.048\dg $\pm$ 0.009\dg, a value just below the significant extension threshold. The upper limit at the 95\% confidence level on the Gaussian extent is $\sigma =$ 0.067\dg. Taking a spectral extraction radius \rspec = 0.15\dg, the spectrum is represented by a power law with a spectral index $\Gamma_{\rm{TeV}} = 2.85 \pm 0.22$ and an integrated energy flux $F_{\rm{1 - 10 \hspace{0.1cm} TeV}} = (1.43 \pm 0.34) \times 10^{-12}$ \dfu. The source is associated with an extended nonthermal X-ray emission (Suzaku~J1427$-$6051, $\sigma = 0.9' \pm 0.1'$) with an energy flux $F_{\rm{X}} = 8.9_{-2.0}^{+3.6} \times 10^{-13}$ \dfu and a spectral index $\Gamma_{\rm{X}} = 3.1_{-0.5}^{+0.6}$ \citep{Fujinaga:2013}. Such a soft spectrum indicates that the photon cutoff energy is likely below the \suzaku energy band. The derived column density $N_{\rm{H}} = 1.1_{-0.25}^{+0.29} \times 10^{23}$ cm$^{-2}$ points toward a possible large distance that could explain why it appears as point-like with H.E.S.S. No associated X-ray point source with similar column density was reported using \xmm data \citep{Fujinaga:2013}. Above 3 GeV, a \fermi point source was detected at the center of HESS~J1427$-$608 with a hard spectral index $\Gamma_{\rm{GeV}} = 1.87 \pm 0.17$ and an integrated flux of $\Phi_{\rm{GeV}} = (3.06$ $\pm$ $0.73)  \times 10^{-10}$ photon cm$^{-2}$ s$^{-1}$ \citep{Guo:2017}. Figure~\ref{fig:HESSJ1427_maps} (left) shows the HGPS significance map with the associated X-ray and GeV sources overlaid. The extended X-ray emission and the hard spectral shape of the \fermi source, both spatially coincident with the TeV emission, make the leptonic scenario plausible as the origin of the TeV emission.

\subsubsection{Multiwavelength constraints}\label{sec:1427_MWL}

We extracted archival radio continuum data toward the region of HESS~J1427$-$608. Figure~\ref{fig:HESSJ1427_maps} (right) shows the MGPS-2 map with the extended nonthermal emission detected by \suzaku overlaid. Two MGPS-2 sources (MGPS~J142755$-$605038 and MGPS~J142714$-$604755) are located inside the \rspec of HESS~J1427$-$608 and a \hii region (reported in the PMN catalog) lies in the northwest (outside) of the H.E.S.S. source extent. The compact source MGPS~J142755$-$605038 ($F = 34.5 \pm 6.4$ mJy) was considered as the possible radio counterpart by \cite{Vorster:2013}. While the authors could reproduce the X-ray and TeV data with a leptonic model implying $B$ = 4 $\mu$G, the predicted radio flux was significantly larger than that of MGPS~J142755$-$605038. On the other hand, they could explain the radio and TeV data with $B$ = 0.4 $\mu$G, but the model fails to account for the flux measured with \suzaku. We defined the ON region such as it encompasses the brightest part of the emission (located between the two MGPS-2 sources) and MGPS~J142755$-$605038 (lying at the center of the \suzaku source), excluding the compact source MGPS~J142714$-$604755 of unknown origin. The radio fluxes were calculated using MGPS-2, SGPS and PMN data and are reported in Table~\ref{tab:radio}. We found a MGPS-2 source with flux roughly six times higher than that of MGPS~J142755$-$605038, which was considered as the radio counterpart in the modeling of \cite{Vorster:2013}. We also note that the flux upper limit derived with the PMN data is underestimated as a result of the masking of MGPS~J142714$-$604755, which leads to a large number of pixels in the ON region that are not taken into account in the flux calculation. The source masking is then the main limitation of this method when using moderate angular resolution instruments (the PSF of the PMN is $4.9'$, compared to $45''$ with the MGPS-2). The best-fit power law gives a spectral index of $\alpha = -0.38 \pm 0.36$, consistent with nonthermal emission. We also derived the spectral index in the region encompassing only the brightest part of the emission (located between the two MGPS-2 sources) and we found a softer spectrum $\alpha = -0.99 \pm 0.34$ that may be due to contamination from the compact source MGPS~J142714$-$604755. The radio morphology in the ON region is more similar to that of a center-filled PWN than a shell-type SNR, and the radio spectral index is compatible with that expected from PWN emission. The softness of the X-ray spectrum reported by \cite{Fujinaga:2013} indicates a relatively low maximum energy reached by particles, which points toward an ancient PWN scenario. In evolved PWNe, radio and X-ray morphologies can be uncorrelated since they originate from particle populations with different injection and radiative loss timescales. Freshly injected particles produce X-ray emission while the oldest particles have already cooled and diffused away from the pulsar, producing a relic radio PWN. We therefore naturally expect the radio extent of an evolved PWN to be larger than the X-ray extent, which is somewhat at odds with that seen in HESS~J1427$-$608. However, it should be noted that radio interferometers cannot reveal large components as extended emission can be partially suppressed during the data reduction;  for example, the MOST telescope does not detect structures on angular scales larger than 20--30'. If Suzaku~J1427$-$6051 and the nonthermal radio emission are both associated with HESS~J1427$-$608, the latter is then probably affected by this instrumental limitation. The displacement of the centroid of the radio emission from that of the X-ray emission is found in other evolved PWNe such as Vela X and can be explained by different particle populations. This could explain why the brightest part of the radio emission is displaced from the Suzaku~J1427$-$6051 centroid, even though it is coincident with the northern tip of the extended X-ray emission. It could thus be connected to the compact source MGPS~J142714$-$604755 if the latter is assumed to be the pulsar at the origin of this multiwavelength wind nebula.

\begin{table*}[t!]
{\small
\centering
\begin{tabular}{lcccccc}
        \hline
        \hline
& \multicolumn{3}{c}{HESS~J1427--608}   &                               \multicolumn{3}{c}{HESS~J1458--608}
        \\
    &   MGPS-2          &                               SGPS                                    &                                PMN 
    &  MGPS-2           &                               Parkes                                  &                                PMN                                              \\
        \hline
        Frequency (GHz) 
&               0.843           &                               1.4                                 &                            4.85 
&               0.843           &                               2.4                                 &                            4.85                                                    \\
        rms (mJy/beam)                  
&       2.2                 &                      8.0                               &                       21.5
&               1.6, 1.6                    &   75.0, 219.5                              &                           11.7, 17.4                                              \\
        Flux density (mJy)      
&         197.1 $\pm$ 16.2 &      161.3 $\pm$ 27.6                   &          $< 90.7$
&         195.4 $\pm$ 22.5, 270.2 $\pm$ 32.5 &    $< 292.6$, $< 1113.7$    &     90.7 $\pm$ 27.2, 164.7 $\pm$ 54.8\\                             
       \hline
\end{tabular}
}
\caption{Estimated radio fluxes toward HESS~J1427$-$608 and HESS~J1458$-$608 with the associated statistical errors. The flux upper limits are given at the 3$\sigma$ confidence level. For HESS~J1458$-$608, the values correspond to those derived within the regions $R_{\rm{ON1}}$ and $R_{\rm{ON2}}$ shown in Figure \ref{fig:HESSJ1458_maps} (right).}
\label{tab:radio}
\end{table*}
\begin{figure*}[th!]
\centering
\includegraphics[scale=0.34]{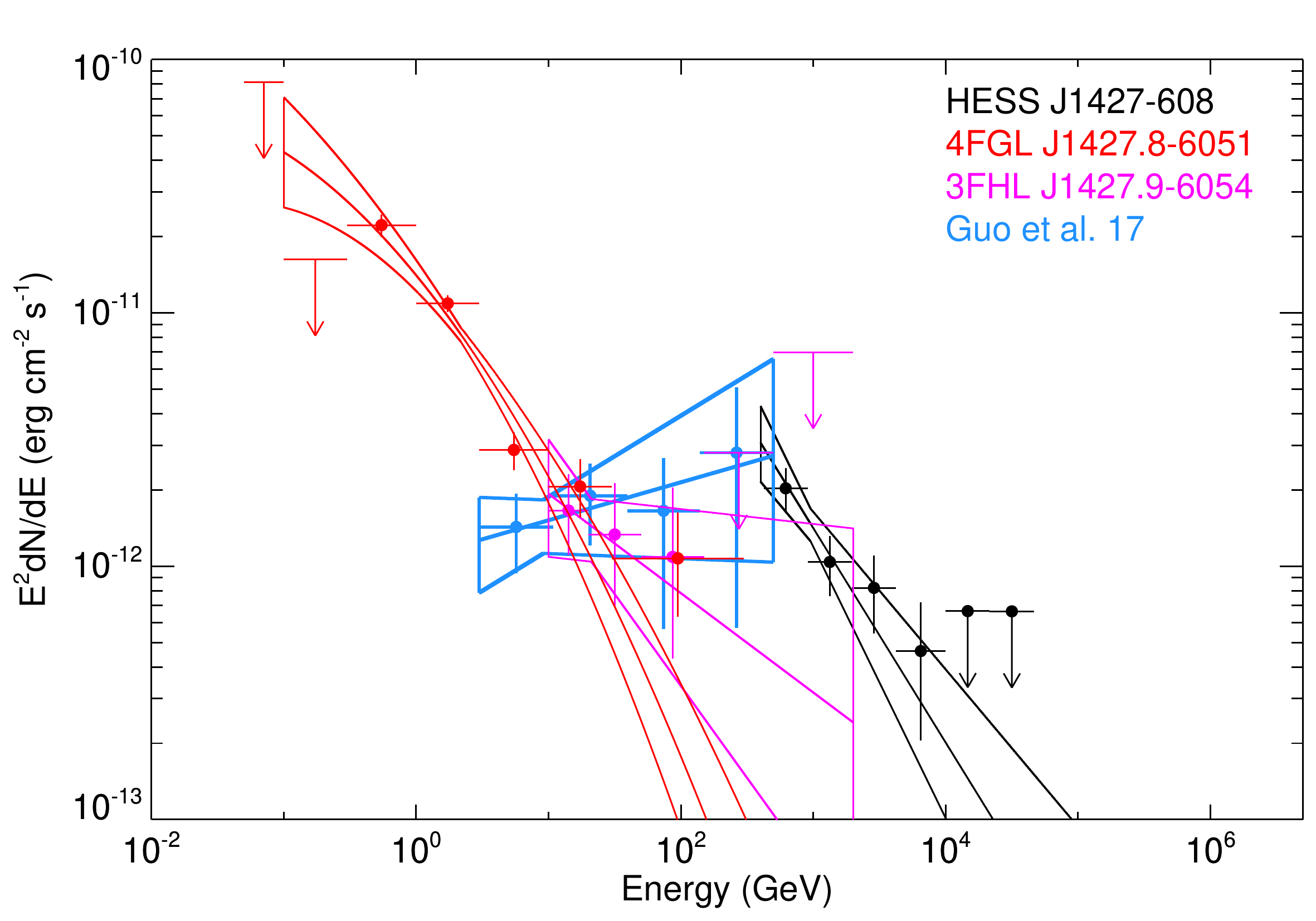} 
\includegraphics[scale=0.35]{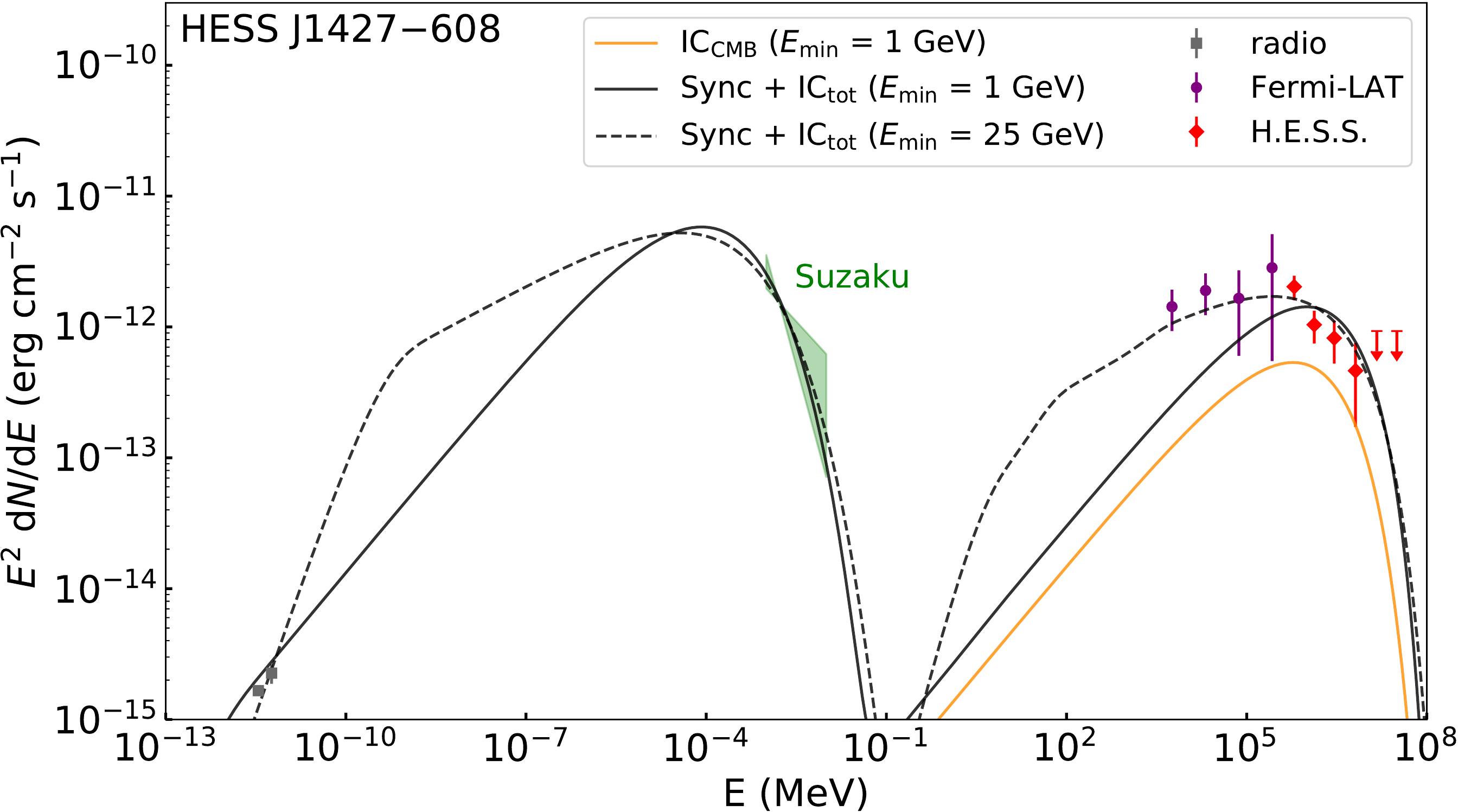}
\caption{(Left) Spectral energy distribution of HESS~J1427$-$608 with those of the associated \fermi sources. The \fermi upper limits are calculated with a confidence level of 95\%. (Right) Broadband nonthermal spectrum of HESS~J1427$-$608 in a leptonic scenario. The solid black line corresponds to the model with $E_{\rm{min}}$ = 1 GeV, $E_{\rm{cut}}$ = 12.3 TeV, $p$ = 1.9, $B$ = 10.1 $\mu$G, and $W_{e} = 4.9\times10^{47}\times \text{(d/8 kpc)}^2$ erg. The model with $E_{\rm{min}}$ = 25 GeV, $E_{\rm{cut}}$ = 19.1 TeV, $p$ = 2.5, $B$ = 9.9 $\mu$G, and $W_{e} = 2.2\times10^{48}\times\text{(d/8 kpc)}^2$ erg is represented by the dashed black line.}
\label{fig:SEDs_HESSJ1427}
\end{figure*}

The ROSAT/PSPC significance map shows no emission toward HESS~J1427$-$608. The X-ray flux upper limit is therefore not constraining ($F_{\rm{X}} < 1.29 \times 10^{-8}$ erg cm$^{-2}$ s$^{-1}$). This is consistent with the high column density value derived in \cite{Fujinaga:2013}, which leads to a large absorption of soft X-ray photons along the line of sight. Thus, we derived a loose constraint on the mean magnetic field, that is, $B < 263.4$ $\mu$G at the 5$\sigma$ confidence level with $N_{\rm{H}} < 1.39 \times 10^{23}$ cm$^{-2}$ \citep{Fujinaga:2013} and $\Gamma_{\rm{TeV}} < 3.7$. Assuming that Suzaku~J1427$-$6051 is the X-ray counterpart of HESS~J1427$-$608, accelerated electrons can be considered in the Thompson regime ($\Gamma_{\rm{X}} \approx \Gamma_{\rm{TeV}}$, with $\Gamma_{\rm{X}}=3.1_{-0.5}^{+0.6}$ and $\Gamma_{\rm{TeV}}= 2.85 \pm 0.22$) and $B$ amounts to $10.9_{-3.2}^{+4.5}$ $\mu$G (from Equation~\ref{eq:B_field}), which is compatible with that obtained for PWNe and SNRs. The latest \xmm source catalog contains 4XMM~J142756.7$-$605214 located near the center of HESS~J1427$-$608 (black cross in Figure~\ref{fig:HESSJ1427_maps}, left), which was not detected in the study of \cite{Fujinaga:2013}. A dedicated X-ray analysis of this region would allow us to determine if 4XMM~J142756.7$-$605214 could be associated with Suzaku~J1427$-$6051.

Three \fermi sources are located inside the \rspec of HESS~J1427$-$608 \citep[3FHL~J1427.9$-$6054, 4FGL~J1427.8$-$6051 and the source detected by][as illustrated in Figure~\ref{fig:HESSJ1427_maps}, left]{Guo:2017}. Figure~\ref{fig:SEDs_HESSJ1427} (left) shows the corresponding spectral energy distributions (SEDs) reported in the \fermi catalogs. The spectrum of the point source detected above 3 GeV \citep{Guo:2017} smoothly connects to the H.E.S.S. measurements while the 4FGL catalog reports a source with a pulsar-like spectrum represented by a logarithmic parabola. If the dominant emission above 3 GeV originates from a PWN powered by a putative pulsar whose emission is reported in the 4FGL catalog, we note that the PWN spectrum is very soft in the TeV range and has a relatively low energy cutoff (in the $\sim$ 500 GeV range), which also points toward an evolved PWN scenario.

\subsubsection{Discussion}\label{sec:discuss_J1427}

Since HESS~J1427$-$608 is unresolved at TeV energies, a blazar or binary origin could have been considered. However, given the existence of an extended X-ray emission spatially coincident with the H.E.S.S. source, these two scenarios are clearly disfavored. With an extended and center-filled nonthermal radio emission detected within HESS~J1427$-$608 and a 4FGL \fermi~cataloged source exhibiting a spectrum reminiscent of a pulsar, our multiwavelength data exploitation strengthens the scenario of an evolved PWN.

To constrain the physical parameters in more detail, we modeled the broadband nonthermal spectrum of HESS~J1427$-$608 using the {\it Naima} package \citep{naima} assuming a one-zone leptonic model. The electron spectrum is in the form d$N_e$/d$E_e$ $\propto$ $E_{e}^{-p}$exp($-E_e/E_{\rm{cut}})$ for $E_e \in$ [$E_{\rm{min}}$, $E_{\rm{max}}$], where $p$ is the particle spectral index. The minimum and maximum energy of the particles are set to $E_{\rm{min}}$ = 1 GeV and $E_{\rm{max}}$ = 1 PeV. Since the distance $d$ is unknown, we arbitrarily set $d$ = 8 kpc \citep[as in][]{Fujinaga:2013}. For the IC scattering, we considered the CMB and the infrared and optical photon fields whose temperature and energy density were estimated using the data from \citet{Porter:2008} included within the \courrier{GALPROP}\footnote{\url{https://galprop.stanford.edu}} project. We obtained $T_{\rm IR}$ = 25 K, $U_{\rm IR}$ = 0.82 eV cm$^{-3}$, $T_{\rm opt}$ = 2005 K, and $U_{\rm opt}$ = 1.07 eV cm$^{-3}$. We used the radio fluxes reported in Table~\ref{tab:radio} (excluding the upper limit derived with PMN data, see Section~\ref{sec:1427_MWL}), the X-ray measurements from \suzaku \citep{Fujinaga:2013}, the GeV data points from \cite{Guo:2017}, and the TeV SED reported in the HGPS. To simultaneously fit the radio and X-ray data, a hard particle spectral index such as $p = 1.9$ is required, and the best fit gives $E_{\rm{cut}} = 12.3_{-1.2}^{+1.2}$ TeV and $B = 10.1_{-0.7}^{+1.0}$ $\mu$G, which is consistent with the magnetic field value inferred in Section~\ref{sec:1427_MWL}. As shown in Figure~\ref{fig:SEDs_HESSJ1427} (right), the low-energy \fermi data points cannot be reproduced with such a hard particle spectral index. Using $p = 2.5$ and $E_{\rm{min}}$ = 25 GeV, the broadband nonthermal emission can be explained with $B = 9.9_{-0.8}^{+0.8}$ $\mu$G and $E_{\rm{cut}} = 19.1_{-1.7}^{+1.7}$ TeV (Figure~\ref{fig:SEDs_HESSJ1427}, right). Although the low-energy cutoff in the electron spectrum is questionable, a value of $E_{\rm{min}}$ = 13 GeV was also required to simultaneously explain the radio and X-ray measurements of the TeV PWN HESS~J1356$-$645 \citep{HESSJ1356_2011}. Such values of $E_{\rm{min}}$ are also in the range of those considered in \cite{Kennel:1984} and \cite{Ackermann_PWN:2011}. We note however that significant radio emission could have been suppressed during the data reduction. Finally, it is worth noting that if HESS~J1427$-$608 is an ancient PWN, a more detailed model is likely needed to explain its broadband nonthermal emission, as this is the case for the Vela X PWN whose multiwavelength modeling requires two electron populations \citep{Hinton:2011}.

Deeper radio observations and dedicated TeV analyses would be helpful to obtain better insight into the source morphology. Pulsation searches on 4FGL~J1427.8$-$6051 are also necessary to possibly assess the nature of this \fermi source as a pulsar and dedicated X-ray observations and analyses are required to search for this putative compact source. If a pulsar is detected with a spin-down luminosity high enough to power a detectable nebula despite its likely large distance, HESS~J1427$-$608 could thus be firmly identified as a TeV PWN. 

\subsection{HESS~J1458$-$608}\label{sec:HESSJ1458}

\begin{figure*}[th!]
    \centering
    \includegraphics[scale=0.43]{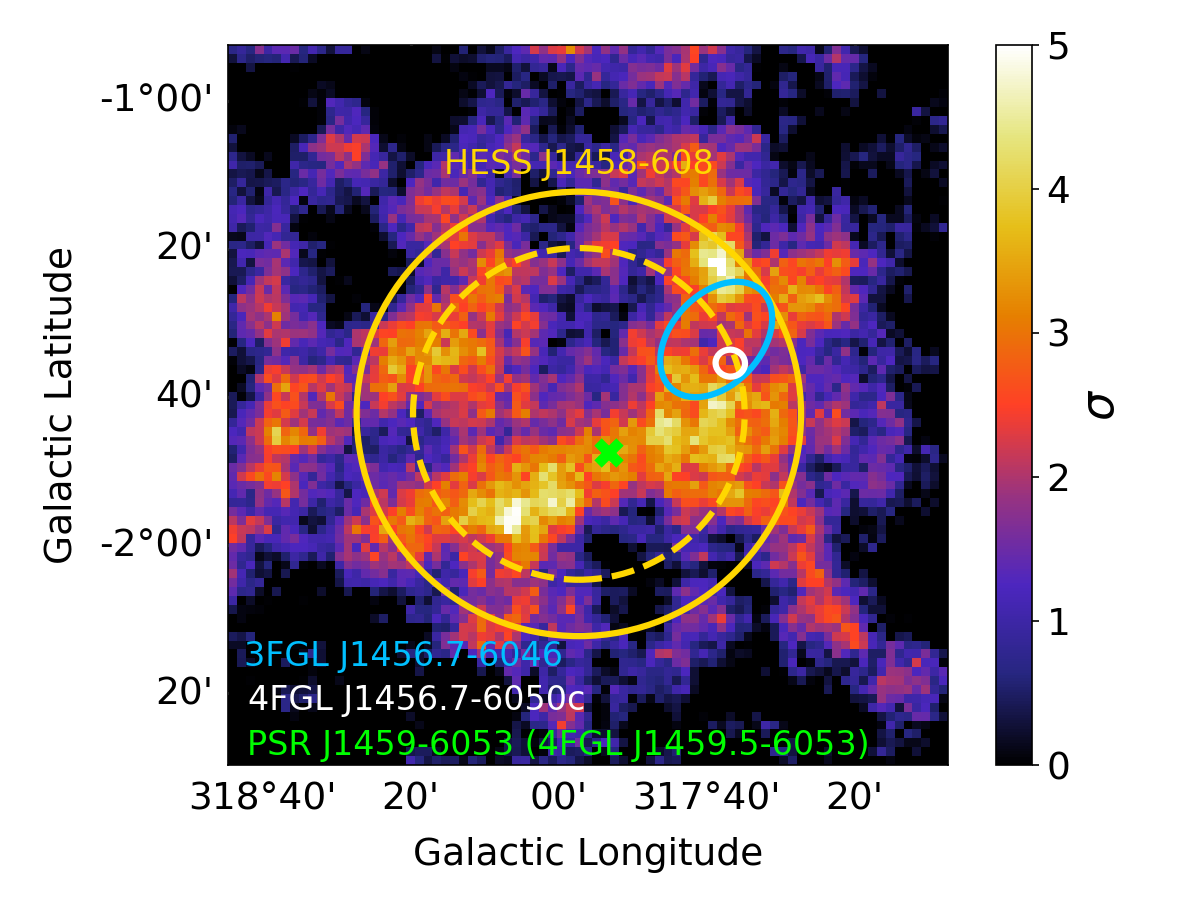}  
    \includegraphics[scale=0.43]{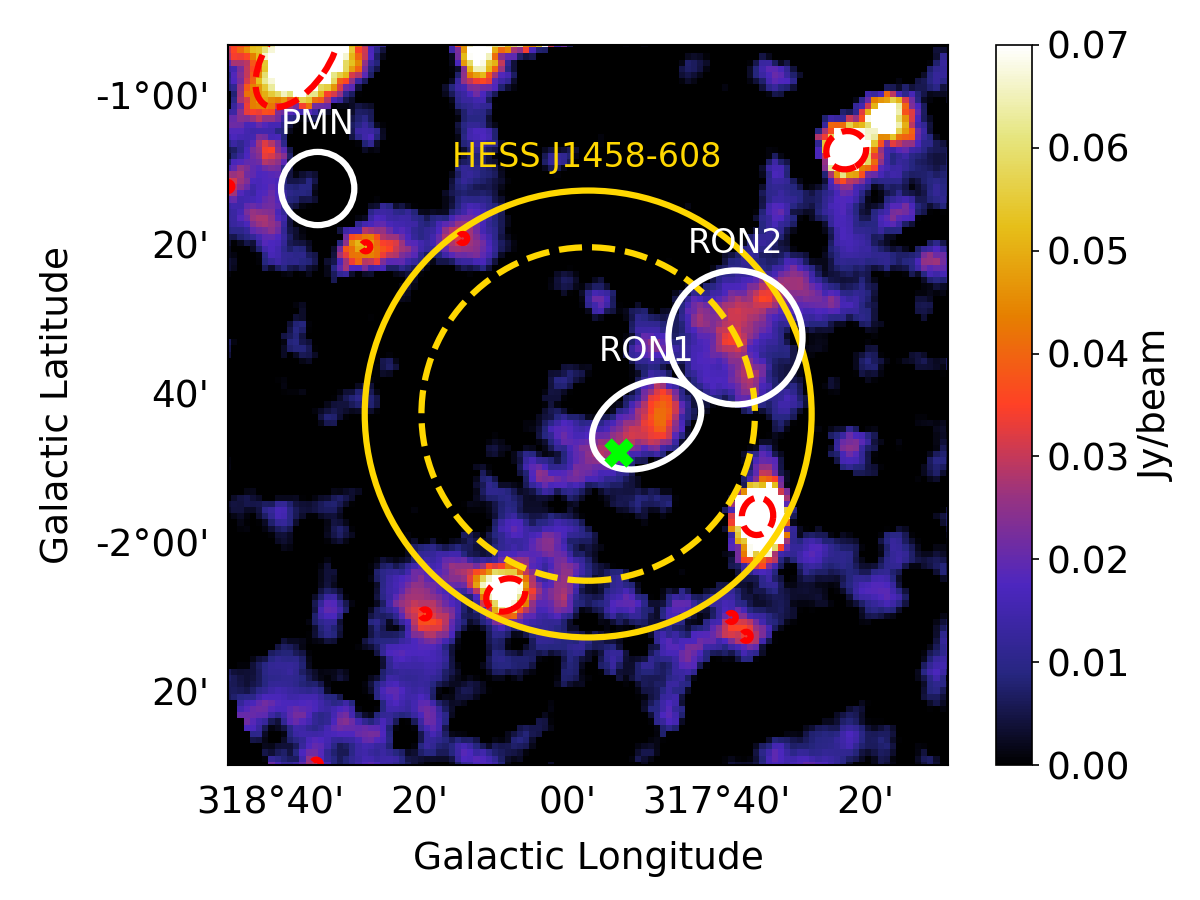}   
    \caption{(Left) HGPS significance map (obtained with a correlation radius of 0.1\dg) centered on HESS~J1458$-$608. The size and spectral extraction radius \rspec of HESS~J1458$-$608 are represented by the dashed and solid yellow circles, respectively. The blue and white circles correspond to the 95\% position uncertainties of the \fermi cataloged sources, while the green cross represents the position of PSR~J1459$-$6053 (associated with 4FGL~J1459.5$-$6053). (Right) PMN map at 4.85 GHz with the radio cataloged sources outside the ON regions represented in red. The flux extraction regions appear in white ($R_{\rm{ON1}}$ and $R_{\rm{ON2}}$) and the FWHM of the PMN is represented at the top left of the image. The yellow circles are the same as those in the left panel.}
    \label{fig:HESSJ1458_maps}
\end{figure*}

\subsubsection{Source presentation}\label{sec:HESSJ1458_presentation}

In the HGPS, HESS~J1458$-$608 is described by a two-dimensional symmetric Gaussian with $\sigma=0.37$\dg $\pm$ 0.03\dg and has a significance of $\sqrt{\rm{TS}}$ = 11.5 \citep{HGPS:2018}. The TeV spectrum is hard with $\Gamma_{\rm{TeV}} = 1.81 \pm 0.14$ and $F_{\rm{1-10 \hspace{0.1cm} TeV}} = (5.28 \pm 1.30) \times 10^{-12}$ \dfu obtained with \rspec = 0.5\dg. At the center of HESS~J1458$-$608 lies a \fermi $\gamma$-ray pulsar PSR~J1459$-$6053 with a spin-down power of $\dot{E} = 9.1 \times 10^{35}$ erg s$^{-1}$ \citep{Ray:2011} and a characteristic age of $\tau_c$ = 64 kyr \citep{Marelli:2011}. Data taken with \swift/XRT (6 ks) revealed a point source, offset from the $\gamma$-ray pulsar by $\sim$ 9.8'' and this source was suggested as the X-ray counterpart of PSR~J1459$-$6053 \citep{Ray:2011}. \xmm observations of 50 ks confirmed the nonthermal X-ray emission from the pulsar with a derived column density of $N_{\rm{H}} = 6.4_{-5.1}^{+8.9} \times 10^{21}$ cm$^{-2}$ \citep{Pancrazi:2012}. Since the pulsar is not detected in the radio band, the distance of the system is unknown. No associated X-ray and $\gamma$-ray emissions originating from a PWN have been detected so far. Figure~\ref{fig:HESSJ1458_maps} (left) shows the H.E.S.S. significance map, indicating that the TeV morphology may be more complex than a symmetric Gaussian and that two \fermi cataloged sources lie in the western part of HESS~J1458$-$608.

\subsubsection{Multiwavelength constraints}\label{sec:mwl_J1458}

We extracted archival radio continuum data that show emission in the vicinity of the pulsar and in the western part of HESS~J1458$-$608, the latter being spatially coincident with the \fermi sources 3FGL~J1456.7$-$6046 and 4FGL~J1456.7$-$6050c (Figure~\ref{fig:HESSJ1458_maps}, right). We defined two regions that encompass the diffuse emissions within the H.E.S.S. source: one in the vicinity of the pulsar (called $R_{\rm{ON1}}$) and one toward the western part of the H.E.S.S. source (called $R_{\rm{ON2}}$). The estimated radio fluxes using MGPS-2, Parkes, and PMN are given in Table~\ref{tab:radio} whose best-fit power law gives $\alpha_{1} = -0.44 \pm 0.18$ (for $R_{\rm{ON1}}$) and $\alpha_{2} = -0.28 \pm 0.20$ (for $R_{\rm{ON2}}$). Both spectral indexes indicate nonthermal radio emission and it is not clear whether these two regions originate from the same system. The spectral indexes of $R_{\rm{ON1}}$ and $R_{\rm{ON2}}$ are compatible with those expected from both PWNe and shell-type SNRs. The spectral index of $R_{\rm{ON1}}$ is even closer to that of a SNR than a PWN, but the absence of a shell-like morphology and the location of the radio emission close to the energetic PSR~J1459$-$6053 makes a PWN scenario more likely.

The pulsar is seen in \chandra, \swift/XRT, \xmm, and \suzaku data but no extended emission appears in these images. The ROSAT/PSPC data show no significant emission and give a tight constraint on the X-ray flux $F_{\rm{X}} < 2.13 \times 10^{-11}$ erg cm$^{-2}$ s$^{-1}$ \citep[with $N_{\rm{H}} < 1.5 \times 10^{22}$ cm$^{-2}$,][and $\Gamma_{\rm{TeV}} < 1.95$]{Pancrazi:2012}, giving $B < 10.1$ $\mu$G at the 5$\sigma$ confidence level.

\begin{figure*}[th!]
    \centering
    \includegraphics[scale=0.34]{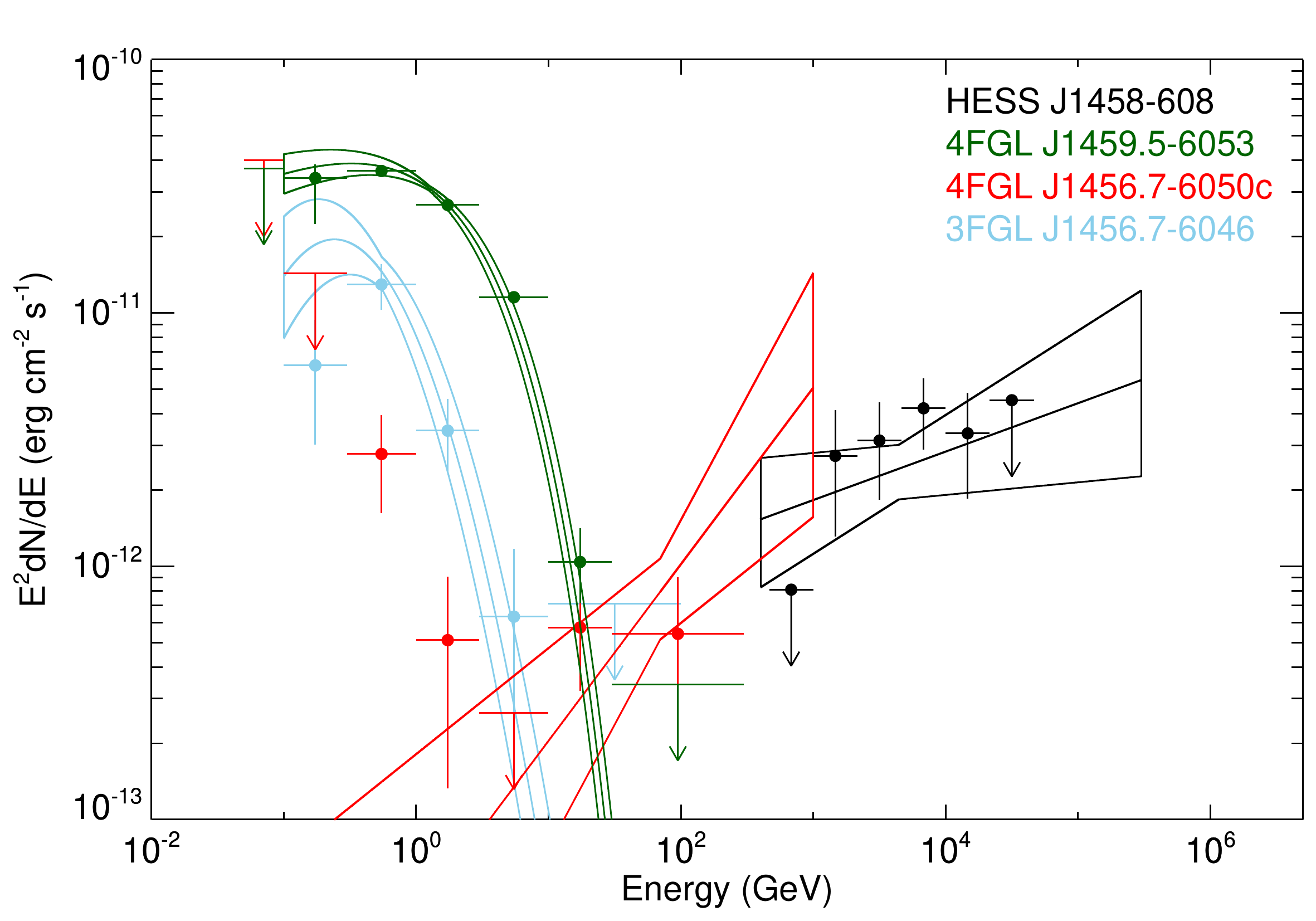}
    \includegraphics[scale=0.36]{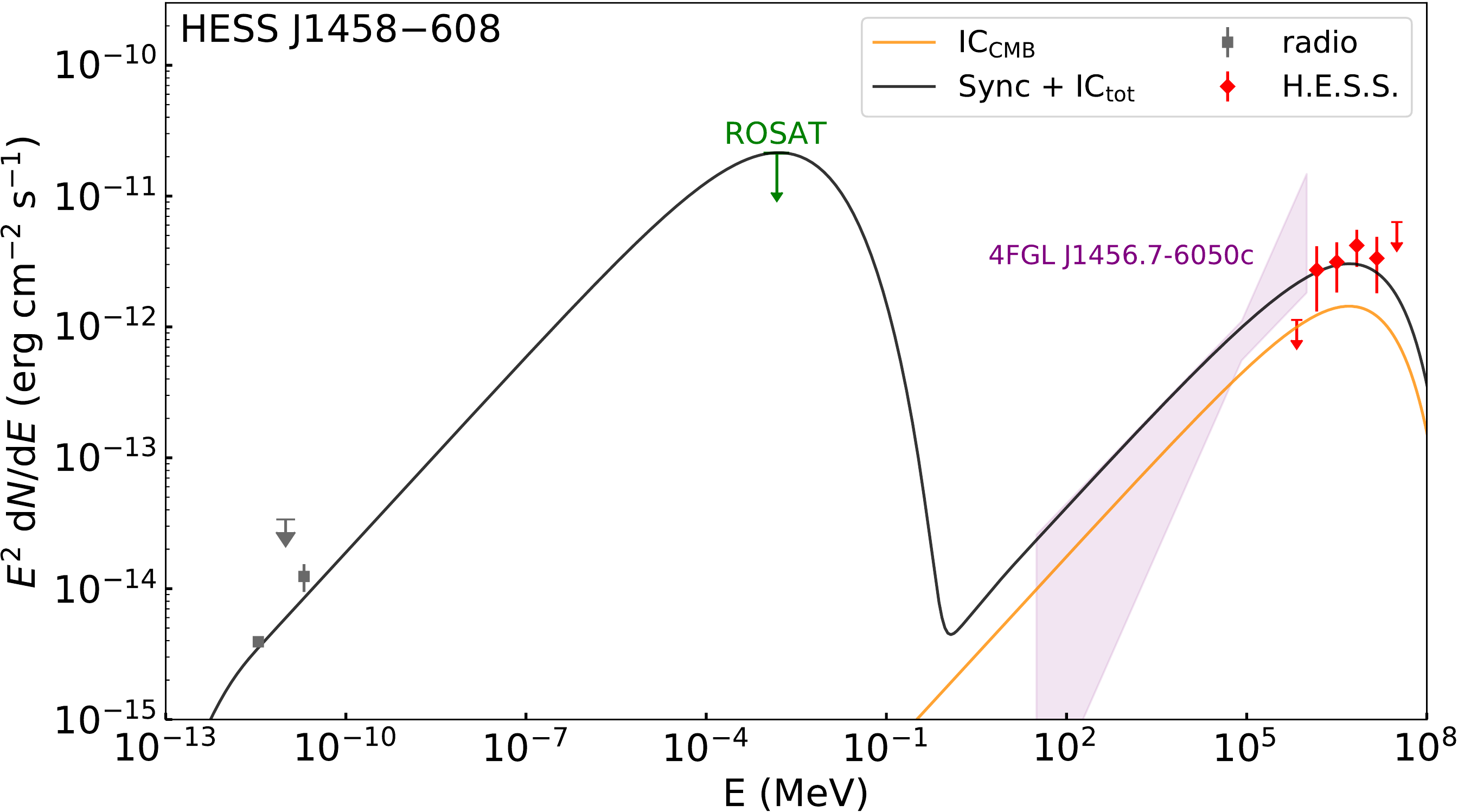}
    \caption{(Left) Spectral energy distribution of HESS~J1458$-$608 with those of the associated \fermi sources. The \fermi upper limits are calculated with a confidence level of 95\%. Since the $\gamma$-ray pulsar PSR~J1459$-$6053 is reported in several \fermi catalogs, only its best-fit spectrum and SED from the 4FGL catalog (4FGL~J1459.5$-$6053) are shown for visibility purpose. (Right) Broadband nonthermal emission of HESS~J1458$-$608 in a leptonic scenario with $E_{\rm{min}}$ = 1 GeV, $E_{\rm{cut}}$ = 60 TeV, $B$ = 10 $\mu$G, $p$ = 2, and $W_{\rm{e}} = 7 \times 10^{47}\times$($d$/8 kpc)$^2$ erg.}
    \label{fig:SEDs_J1458}
\end{figure*}

The pulsar PSR~J1459$-$6053 is reported in the 2PC, 3FGL, 3FHL, and 4FGL catalogs. The \fermi catalogs also contain two sources located in the western part of HESS~J1458$-$608 (3FGL~J1456.7$-$6046 and 4FGL~J1456.7$-$6050c, the latter being labeled as confused). Figure~\ref{fig:SEDs_J1458} (left) shows that the best-fit spectrum of 4FGL~J1456.7$-$6050c in the whole energy range is hard ($\Gamma_{\rm{GeV}} = 1.30 \pm 0.28$) and connected to the spectrum of HESS~J1458$-$608. The SED at low energy of 4FGL~J1456.7$-$6050c is likely contaminated by the nearby $\gamma$-ray pulsar, as might be the case for the SED of 3FGL~J1456.7$-$6046. However, we keep in mind that the detection of 4FGL~J1456.7$-$6050c may be spurious since the source is labeled as confused.

\subsubsection{Discussion}

HESS~J1458$-$608 is a puzzling source, with a complex and elongated TeV morphology, and shelters the energetic X-ray and $\gamma$-ray pulsar PSR~J1459$-$6053. With nonthermal radio emission located in the western part of the H.E.S.S. source, which is spatially coincident with a GeV source exhibiting a hard spectrum (4FGL~J1456.7$-$6050c), part of the emission from HESS~J1458$-$608 could originate from a PWN. We also reported nonthermal radio emission in the vicinity of PSR~J1459$-$6053. These extended nonthermal radio emissions, possibly connected to each other, are spatially coincident with the western part of the elongated shape of HESS~J1458$-$608, as seen in Figure~\ref{fig:HESSJ1458_maps}.

As for HESS~J1427$-$608, we modeled the broadband nonthermal spectrum of HESS~J1458$-$608 assuming a one-zone leptonic scenario and a distance of $d = 8$ kpc. The electron spectrum is defined as in Section~\ref{sec:discuss_J1427} and the temperature and energy density of the infrared and optical photon fields are $T_{\rm{IR}} = 25$ K, $U_{\rm{IR}} = 0.88$ eV cm$^{-3}$, $T_{\rm{opt}} = 2005$ K and $U_{\rm{opt}} = 1.94$ eV cm$^{-3}$ (estimated using GALPROP code). We used the radio fluxes (from both regions) listed in Table~\ref{tab:radio}, the upper limit on the X-ray flux derived in Section~\ref{sec:mwl_J1458} and the TeV SED from the HGPS. Owing to the lack of X-ray data points and the absence of a VHE cutoff, $B$ and $E_{\rm{cut}}$ cannot be accurately derived. As shown in Figure~\ref{fig:SEDs_J1458} (right), the data can be reproduced with $E_{\rm{min}}$ = 1 GeV, $E_{\rm{cut}}$ = 60 TeV, $B$ = 10 $\mu$G, $p$ = 2 and $W_{\rm{e}} = 7 \times 10^{47}$ $\times$($d$/8 kpc)$^2$ erg. For illustration, the best-fit spectrum of 4FGL~J1456.7$-$6050c is represented in Figure~\ref{fig:SEDs_J1458} (right), keeping in mind that it was derived for a point-source morphology and that the source detection needs to be confirmed.

Deeper X-ray observations and dedicated TeV analyses are required to provide better insight into the source morphology and to better constrain the broadband nonthermal spectrum. A study at HE would also be useful to confirm the detection of a source (and to unveil its morphology) exhibiting a hard spectrum connected to that of HESS~J1458$-$608. \\

\subsection{HESS~J1427$-$608 and HESS~J1458$-$608 within the context of the population of TeV PWNe}\label{sec:PWNpop}

We found extended nonthermal radio emission toward HESS~J1427$-$608 and HESS~J1458$-$608. For both sources, the faint radio emission, although not clearly correlated with the TeV morphology, points toward an ancient PWN. The mean magnetic field (inferred from Equation~\ref{eq:B_field} and in more detail through broadband modeling) is consistent with such a scenario and we found a pulsar-like and a PWN-like spectrum in the vicinity of HESS~J1427$-$608 and HESS~J1458$-$608. We note however that HESS~J1458$-$608 could be a composition of multiple sources, since the hard spectrum seen by the \fermi (although this detection needs to be confirmed) originates from the western edge of the TeV source with a nonthermal radio counterpart. Therefore, it is not clear whether this part is connected to the main emission region of HESS~J1458$-$608, where the pulsar PSR~J1459$-$6053 is located. Assuming a PWN scenario for HESS~J1427$-$608 and HESS~J1458$-$608, we want to estimate the pulsar spin-down power and its characteristic age to place both sources within the context of the population of TeV PWNe.

The H.E.S.S. collaboration undertook a PWN population study \citep[using a total of 37 sources including firmly identified PWNe and PWN candidates,][]{PWNpop:2018} and derived a relation between the TeV surface brightness $S$ and the pulsar spin-down power $\dot{E}$ as follows:
\begin{equation}\label{eq:scalinglaw}
S = 10^{(30.62 \pm 0.13)} \times \dot{E}^{0.81 \pm 0.14}
,\end{equation}
where $S$ and $\dot{E}$ are in units of erg pc$^{-2}$ s$^{-1}$ and 10$^{36}$ erg s$^{-1}$, respectively. The surface brightness can be written as
\begin{equation}
S = \frac{L_{\rm{1-10 \hspace{0.1cm} TeV}}}{4 \pi R_{\rm{PWN}}^2} \approx  \frac{F_{\rm{1-10 \hspace{0.1cm} TeV}}}{\sigma^2}
,\end{equation}
where $L_{\rm{1-10 \hspace{0.1cm} TeV}}$, $F_{\rm{1-10 \hspace{0.1cm} TeV}}$, $R_{\rm{PWN}}$ and $\sigma$ are, respectively, the luminosity (erg s$^{-1}$) and the integral energy flux (erg pc$^{-2}$ s$^{-1}$) between 1 and 10 TeV, the radius of the PWN (pc) and the Gaussian $\sigma$ extent (radians). Since no pulsar was detected in the vicinity of HESS~J1427$-$608 and the distance of PSR~J1459$-$6053 is unknown, HESS~J1427$-$608 and HESS~J1458$-$608 were not considered in the H.E.S.S. PWN population study. For HESS~J1427$-$608, we have $F_{\rm{1 - 10 \hspace{0.1cm} TeV}} = (1.43 \pm 0.34) \times 10^{-12}$ \dfu and $\sigma$ < 0.067\dg, which gives $\dot{E} >$ 1.38 $\times$ $10^{36}$ erg s$^{-1}$. Assuming that Suzaku~J1427$-$6051 is the X-ray counterpart of HESS~J1427$-$608, the TeV extent should be larger than the X-ray extent ($\sigma_{\rm{X}}$ = 0.9') as it  originates from older electrons that have diffused farther away from the pulsar than those emitting X-ray synchrotron. With $\sigma$ > 0.9', we obtained $\dot{E} <$ 6.91 $\times$ 10$^{38}$ erg s$^{-1}$. For HESS~J1458$-$608, with $F_{\rm{1-10 \hspace{0.1cm} TeV}} = (5.28 \pm 1.30) \times 10^{-12}$ \dfu and $\sigma=0.37$\dg $\pm$ 0.03\dg, we obtained $\dot{E} = 2.1_{-1.6}^{+3.4} \times 10^{35}$ erg s$^{-1}$, which is only 2$\sigma$ away from the measurement of $\dot{E}$ = 9.1 $\times$ $10^{35}$ erg s$^{-1}$ for PSR~J1459$-$6053 \citep{Ray:2011}. If PSR~J1459$-$6053 and HESS~J1458$-$608 are associated, this difference could indicate that either the integral energy flux of the TeV source was underestimated or its Gaussian $\sigma$ extent was overestimated. Given the complex TeV morphology of HESS~J1458$-$608 (Figure~\ref{fig:HESSJ1458_maps}, left), possible source confusion could lead to a $\sigma$ extent larger than that associated with the PWN produced by PSR~J1459$-$6053.

The characteristic age of a pulsar $\tau_c$ is expressed as
\begin{equation}
    \tau_c = \frac{(n-1)}{2} \times (t_{\rm{age}} + \tau_0) \hspace{0.5cm} \rm{yr}
,\end{equation}
where $n$ is the braking index, $t_{\rm{age}}$ the age of the system, and $\tau_0$ the initial characteristic age of the pulsar \citep{Gaensler_Slane:2006}. Assuming $n = 3$ (appropriated for magnetic dipoles), we have $\tau_c \approx t_{\rm{age}}$ for evolved systems ($t_{\rm{age}} >> \tau_0$). Since these relic PWNe no longer inject particles into the system, the maximum energy reached by particles is limited by radiative losses. The break energy $E_{\rm{b}}$ (above which electrons significantly suffer from synchrotron losses) is found equating $t_{\rm{age}}$ = $\tau_{\rm{sync}}$, where $\tau_{\rm{sync}}$ is the synchrotron loss time
\begin{equation}
    \tau_{\rm{sync}} = (1.25 \times 10^{3}) \times E_{\rm{TeV}}^{-1} B_{\rm{100 \mu G}}^{-2} \hspace{0.5cm} \rm{yr}
    \label{eq:sync}
,\end{equation}
where $E_{\rm{TeV}}$ and $B_{\rm{100 \mu G}}$ are the particle energy and mean magnetic field in units of TeV and 100 $\mu$G, respectively \citep{Parizot:2006}. In evolved systems limited by radiative losses, we can therefore assume $E_{\rm{b}} = E_{\rm{cut}}$. Our estimate on $\tau_c$ relies on the assumption that HESS~J1427$-$608 and HESS~J1458$-$608 are evolved PWNe and that we assume $\tau_c = t_{\rm{age}}$. For HESS~J1427$-$608, we used $B = 10.0_{-0.9}^{+1.1}$ $\mu$G and $E_{\rm{cut}} = 15.7_{-4.6}^{+5.1}$ TeV (combining the constraints obtained with $E_{\rm min} = 1$ and 25 GeV in Section~\ref{sec:discuss_J1427}) and we deduced an age $\tau_c = t_{\rm{age}}$ = $8.0$ $[4.9, 13.6]$ kyr (Equation~\ref{eq:sync}). For HESS~J1458$-$608, we derived a lower limit of $E_{\rm{cut}} > 30$ TeV since no break or cutoff is visible in Figure~\ref{fig:SEDs_J1458} (right). In order to bound the cutoff energy and the mean magnetic field, we assumed $E_{\rm{cut}} < 100$ TeV and $B > $ 3 $\mu$G, which are appropriate for evolved PWNe. With $B < $ 10.1 $\mu$G (derived in Section~\ref{sec:mwl_J1458}), we found $1.2 < \tau_c = t_{\rm{age}} < 46.3$ kyr.

\begin{figure}[t!]
    \centering
    \includegraphics[scale=0.39]{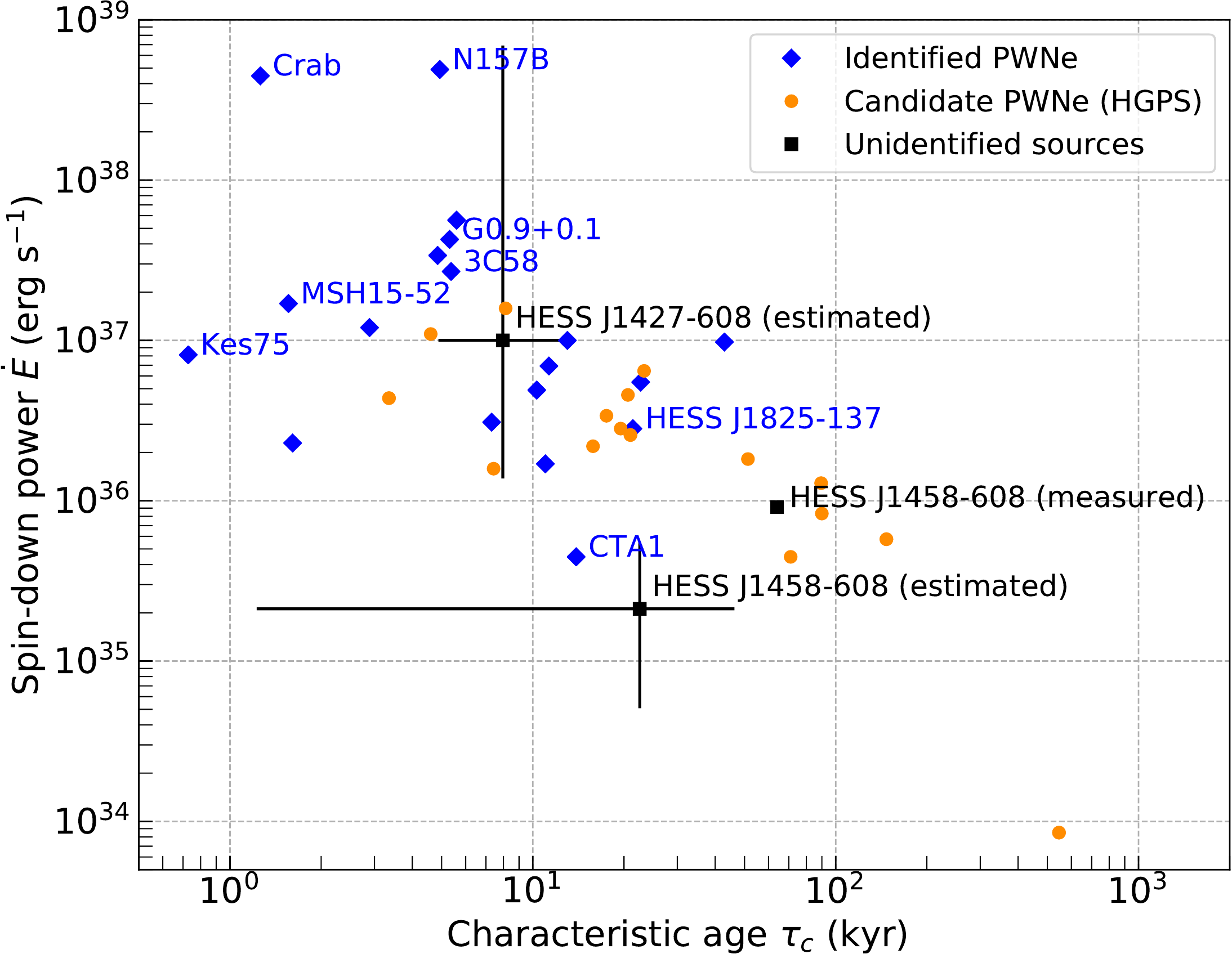}
    \caption{Spin-down power $\dot{E}$ and characteristic age $\tau_c$ of pulsars with either a firmly identified TeV PWN (blue diamonds) or a PWN candidate (orange circles). Assuming a PWN scenario, the values obtained for the unidentified sources HESS~J1427$-$608 and HESS~J1458$-$608 (black squares, labeled as \gui{estimated}) are calculated (see Section~\ref{sec:PWNpop} for details). The measured values of $\dot{E}$ and $\tau_c$ for the pulsar PSR~J1459$-$6053 \citep{Ray:2011, Marelli:2011}, lying at the center of HESS~J1458$-$608, are also reported (labeled as \gui{measured}). The figure is reproduced from \cite{PWNpop:2018} with the addition of HESS~J1427$-$608 and HESS~J1458$-$608.}
    \label{fig:Edot_tauc}
\end{figure}

\begin{figure*}[t!]
    \centering
    \includegraphics[scale=0.43]{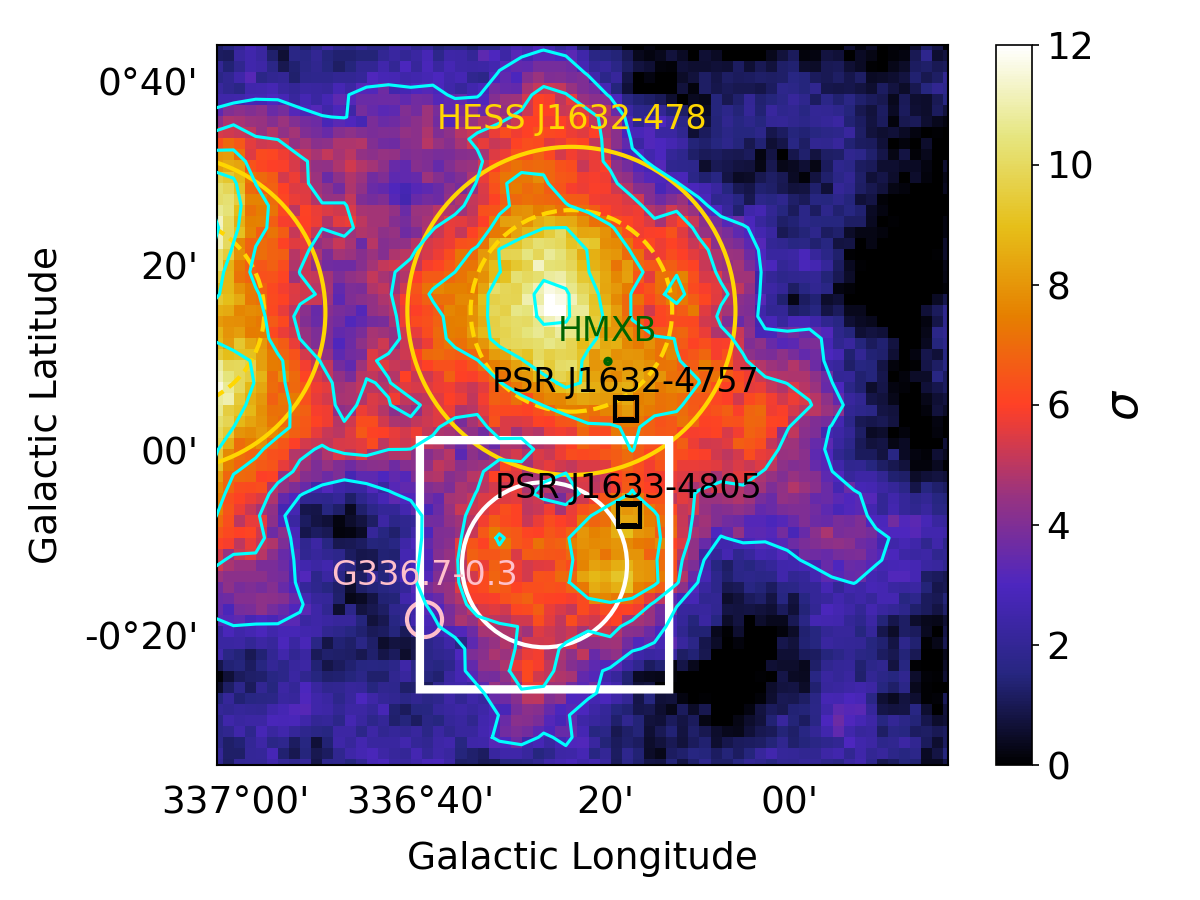}
    \includegraphics[scale=0.43]{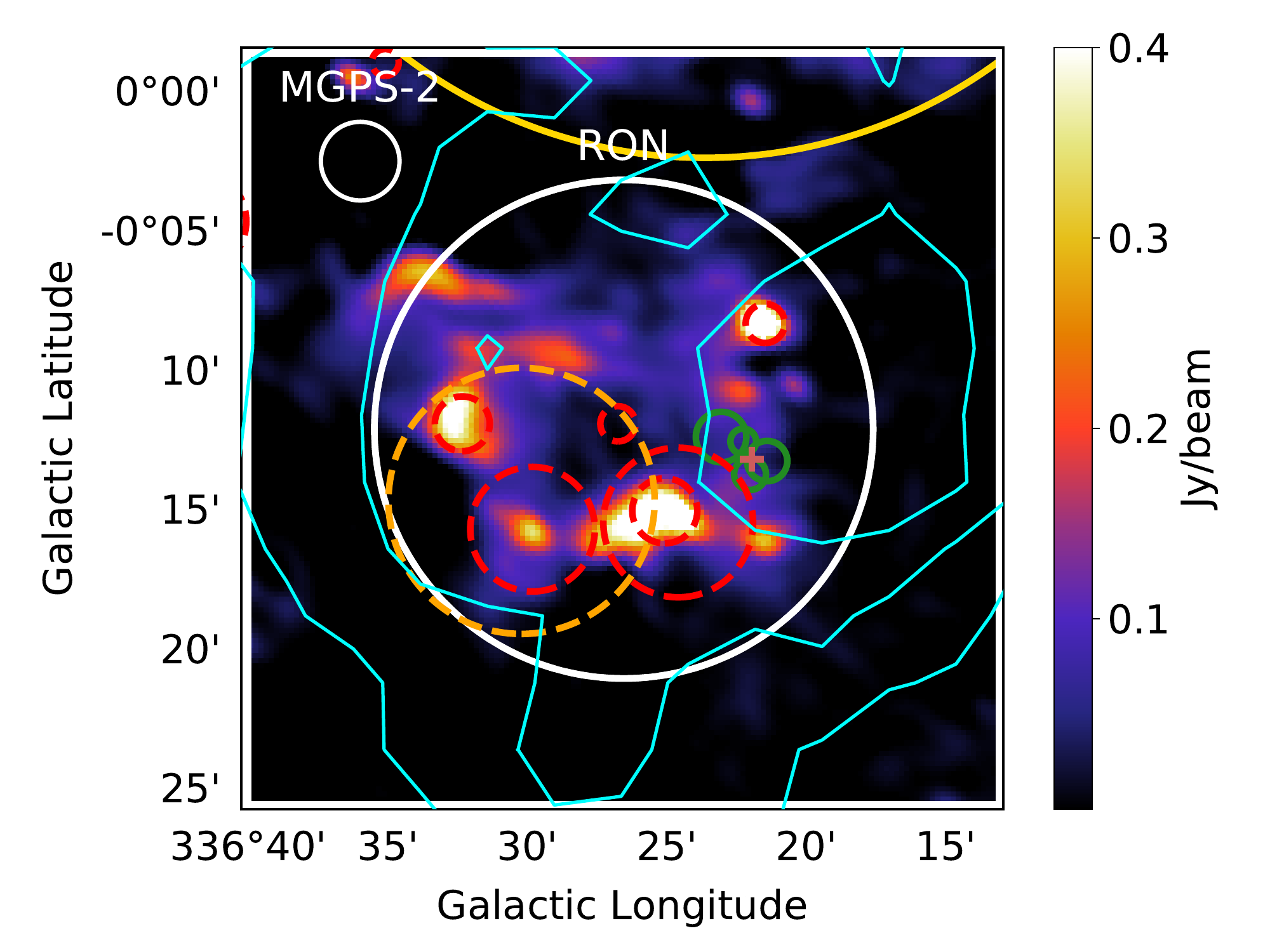}
    \caption{(Left) HGPS significance map (with a correlation radius of 0.1\dg) toward the region of HESS~J1632$-$478. The dashed and solid yellow circles indicate the Gaussian $\sigma$ extent and the \rspec of HESS~J1632$-$478, respectively. A high-mass X-ray binary (HMXB) and the pulsar PSR~J1632$-$4757 are located within the \rspec. The H.E.S.S. contours from 3$\sigma$  to 11$\sigma$ appear in cyan. The $\gamma$-ray excess studied in this work is located within the white box (with a peak significance of $\sim$ 8), near the pulsar PSR~J1633$-$4805 and the uncertain SNR G336.7$-$0.3. The white circle represents the radio flux extraction region $R_{\rm{ON}}$. (Right) MGPS-2 map at 843 MHz of the region within the white box seen in the left panel. The instrument beam is given on the top left corner. The dashed orange and red circles correspond to the cataloged MC \citep{Rice:2016} and \hii regions \citep{Anderson_WISE:2014}, respectively. The pink cross represents a source reported in the Galactic O star catalog \citep{GOSC:2013}, while the green circles indicate the $\sigma$ extent (containing the PSF) of the \xmm extended sources \citep{Webb:2020}. The contours and solid circles are the same as in the left panel.}
    \label{fig:HESSJ1632}
\end{figure*}

Figure~\ref{fig:Edot_tauc} depicts the spin-down power and the characteristic age of pulsars with either a firmly identified TeV PWN or a PWN candidate \citep[reproduced from][with the addition of HESS~J1427$-$608 and HESS~J1458$-$608]{PWNpop:2018}. The estimated values for HESS~J1427$-$608 and HESS~J1458$-$608 are contained within those of the sample of identified PWNe and PWN candidates. However, HESS~J1427$-$608 and HESS~J1458$-$608 appear to be similar to a relatively young PWN (with a high $\dot{E}$) and an evolved PWN (with an $\dot{E}$ estimate lying among the lowest values of the TeV PWN population), respectively, whereas their $\gamma$-ray spectrum (shown in Figure~\ref{fig:SEDs_HESSJ1427} and~\ref{fig:SEDs_J1458}) may suggest the opposite. It is nevertheless possible that the extent of HESS~J1458$-$608 was overestimated as a result of source confusion and that the hard TeV spectral shape originates from multiple source contribution. Owing to the inherent uncertainties in such approach, we cannot draw precise conclusions except that HESS~J1427$-$608 and HESS~J1458$-$608 seem to broadly follow the general trend of the population of TeV PWNe.

Finally, we note that the characteristic age of HESS~J1427$-$608 is comparable to that of the evolved PWN HESS~J1356$-$645 ($\tau_c$ = 7.3 kyr). Assuming that the intrinsic size of HESS~J1427$-$608 is similar to that of HESS~J1356$-$645 ($R \approx$ 8.4$d_{2.4}$ pc, where $d_{2.4}$ is the distance in units of 2.4 kpc), the distance to HESS~J1427$-$608 would be $d \approx$ 10 kpc using an angular TeV size of $\sigma$ = 0.048\dg (or $d$ > 7 kpc for $\sigma$ < 0.067\dg). HESS~J1458$-$608 would be more comparable to the PWN HESS~J1825$-$137 ($\tau_c$ = 21 kyr, $R \approx$ 50$d_{4}$ pc). Assuming that their intrinsic sizes are similar, that would place HESS~J1458$-$608 at a distance $d \approx 8$ kpc for an angular TeV size of $\sigma$ = 0.37\dg. The distances to HESS~J1427$-$608 and HESS~J1458$-$608 could therefore be compatible with the relatively large column densities derived for their associated X-ray sources (Suzaku~J1427$-$6051 and PSR~J1459$-$6053, see Sections~\ref{sec:HESSJ1427_presentation} and~\ref{sec:HESSJ1458_presentation}). Regardless of the dominant particle transport mechanism \citep[diffusion or advection, as discussed for HESS~J1825$-$137 in][]{1825:2019}, HESS~J1427$-$608 and HESS~J1458$-$608 could share similar characteristics with the well-known and bright TeV PWNe HESS~J1356$-$645 and HESS~J1825$-$137, respectively.

\section{New VHE $\gamma$-ray source candidates}\label{sec:targets_CTA}

Although very robust, the HGPS detection pipeline, relying on an iterative template fitting in which the $\gamma$-ray sources are treated as two-dimensional symmetric Gaussians, faced the difficulty of revealing multiple components in regions with large source confusion. The visual inspection of the HGPS images led us to focus on two possible $\gamma$-ray excesses lying in such complex regions, which eluded detection by the algorithm. The first is located south of the unidentified source HESS~J1632$-$478 and the other lies at the position of the synchrotron-emitting SNR G28.6$-$0.1, at the edge of HESS~J1843$-$033. Recently, \citet{Remy:2020} developed an alternative algorithm based on image processing and pattern recognition techniques to detect $\gamma$-ray sources without strong prior morphological assumptions: structural information is extracted using an edge detection operator and the detected objects are found as local maxima after applying the Hough circle transform. This algorithm, applied on the HGPS maps, has provided a list of objects that warrant further investigation. Each of the two above-mentioned excesses, which we independently identified after examining the HGPS maps, has a counterpart in the catalog of \citet{Remy:2020} as a detected object uncataloged in the HGPS (labeled \texttt{HC\_147} and \texttt{HC\_382}, respectively). Although this strengthens the interest of these two excesses, only a dedicated H.E.S.S. data analysis could confirm these as proper VHE sources. In the following, we consider them as such and investigate their origin through the existing multiwavelength data.

\subsection{The south of HESS~J1632$-$478}

The unidentified source HESS~J1632$-$478 is described in the HGPS by a Gaussian component with $\sigma = 0.18$\dg $\pm$ 0.02\dg. The HGPS significance map is shown in Figure~\ref{fig:HESSJ1632} (left) and shows a $\gamma$-ray excess with a peak significance at about 8$\sigma$ in the south of HESS~J1632$-$478 (outside of the HGPS \rspec). The pulsar PSR~J1633$-$4805 \citep[$\dot{E}$ = 8.5 $\times$ $10^{33}$ erg s$^{-1}$ and $d = 6.2$ kpc,][]{ATNF:2005} and the uncertain SNR G336.7$-$0.3 \citep{Green:2017} are located near this $\gamma$-ray excess. 

We explored archival radio continuum data toward the south of HESS~J1632$-$478. The MGPS-2 map is given in Figure~\ref{fig:HESSJ1632} (right) and shows a bright emission with some distorted shell-like structures. Several \hii regions reported in the WISE catalog \citep{Anderson_WISE:2014} and one MC \citep{Rice:2016} are spatially coincident with the $\gamma$-ray excess. Using data from the MGPS-2, SGPS and Parkes, we derived a radio spectral index of $\alpha = 0.64 \pm 0.04$ in the ON region encompassing the radio emission (white circle in Figure~\ref{fig:HESSJ1632}, right). The spectral index indicates that the radio emission is thermal in nature and could originate from the \hii regions. Since these \hii regions do not constitute the entire radio emission, it is possible that there is a subdominant nonthermal emission. 

We also found a X-ray point-source detected by ROSAT/PSPC (2RXS~J163352.2$-$480643) located in the western part of the ON region, which is also seen by \asca/GIS although not cataloged. Four \xmm extended sources are also reported in this region (see Figure~\ref{fig:HESSJ1632}, right): 4XMM~J163353.3$-$480520 ($\sigma = 54.4''$, containing the PSF of $\sim$ $6''$), 4XMM~J163350.8$-$480601 ($\sigma = 27.7''$), 4XMM~J163354.9$-$480659 ($\sigma = 34.3''$), and 4XMM~J163350.4$-$480707 ($\sigma = 43.4''$). These X-ray sources have unknown origin and the Simbad database does not report any known object at these positions. A source, reported in the Galactic O star catalog \citep{GOSC:2013}, is lying at the center of the four extended \xmm sources, which is spatially coincident with a \swift/XRT cataloged source (2SXPS~J163352.3$-$480640).

The \fermi extended source FGES~J1633.0$-$4746 \citep[described by a disk with a radius $r$ = 0.61\dg;][]{FGES:2017} encompasses the TeV $\gamma$-ray excess but its large extent points toward possible source confusion. The \fermi residual count map above 10 GeV shows emission that is spatially coincident with the TeV $\gamma$-ray excess, but a dedicated analysis would be required to potentially reveal an extended component.

To conclude, we found radio emission that is spatially coincident with the significant $\gamma$-ray excess in the south of HESS~J1632$-$478, exhibiting a complex morphology reminiscent of a distorted shell-like structure with a thermal spectrum. Several \hii regions located within the $\gamma$-ray excess indicate that the TeV emission could arise from a star-forming region, which can accelerate particles up to $\gamma$-ray emitting energies (with colliding stellar winds). Such regions were recently proposed as sources of Galactic CRs \citep{Aharonian:2018}. Extended X-ray sources are found in the western part of this region with unknown origin. Deeper X-ray observations would help understand the nonthermal processes occurring in this region. Detailed HE and VHE $\gamma$-ray data analyses would be of great interest to investigate the morphology and spectrum of this $\gamma$-ray excess and potentially reveal a new VHE source.

\subsection{The SNR G28.6$-$0.1 embedded in HESS~J1843$-$033}

\begin{figure*}[th!]
    \centering
    \includegraphics[scale=0.43]{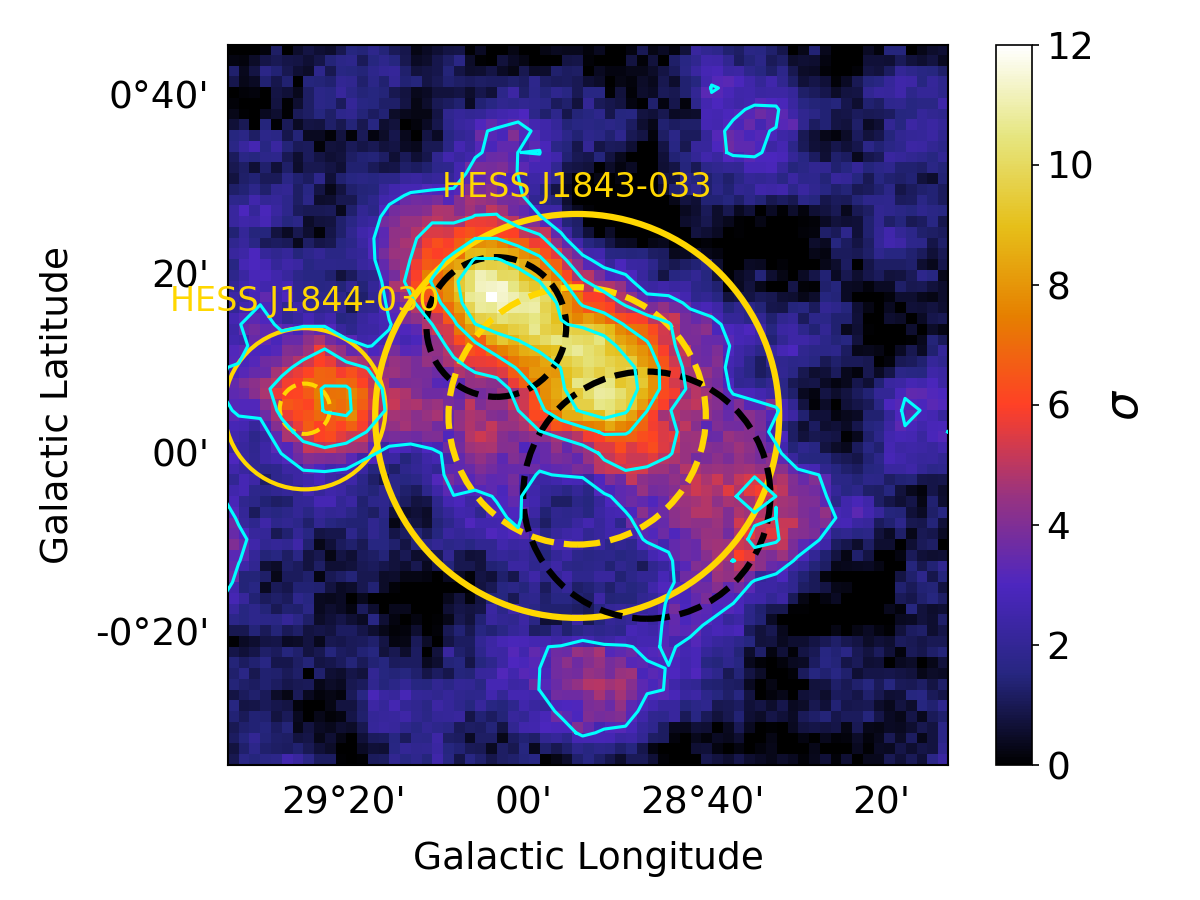}
    \includegraphics[scale=0.43]{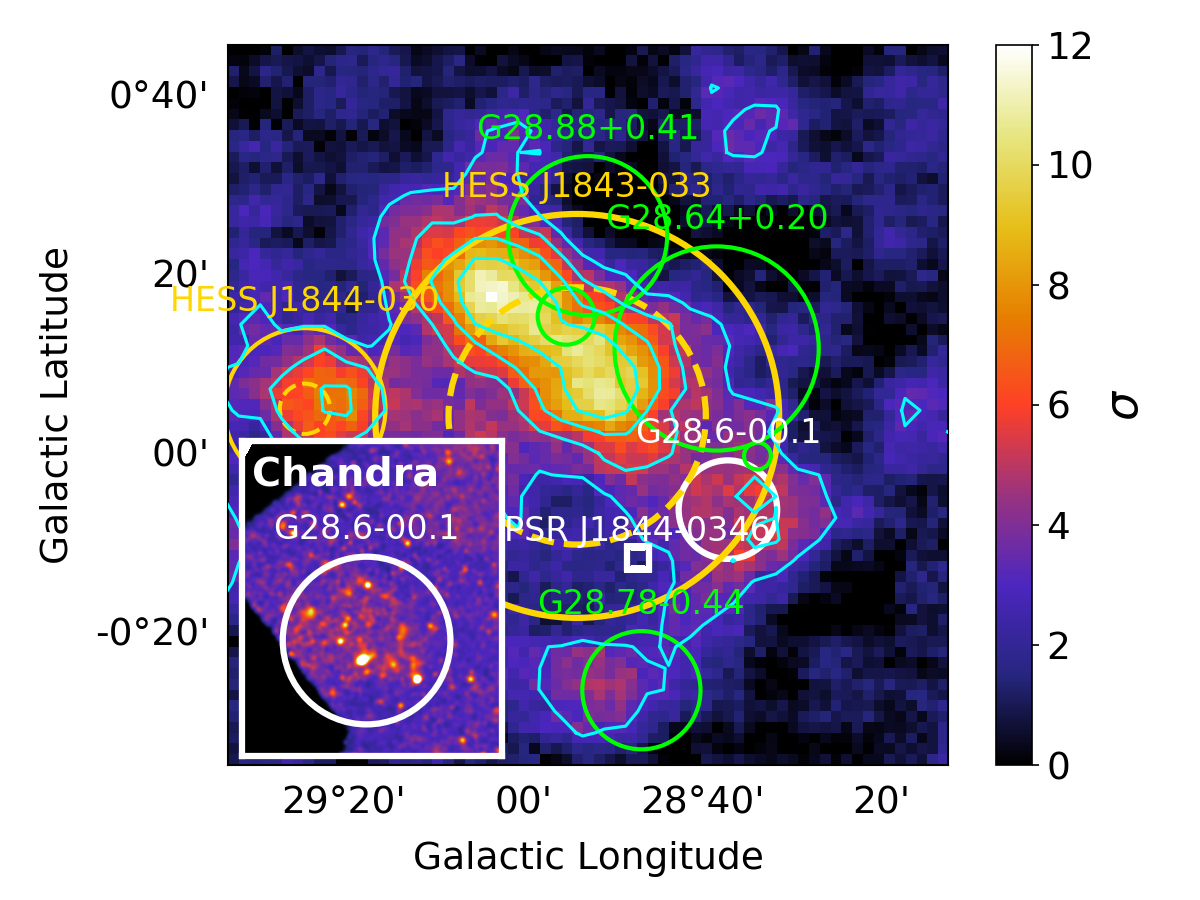}
    \caption{(Left) HGPS significance map of HESS~J1843$-$033 (obtained with a correlation radius of 0.1\dg). The dashed and solid yellow circles indicate the Gaussian $\sigma$ extent and the \rspec. The dashed black circles represent the two Gaussian components previously detected by the HGPS pipeline, which were finally merged into HESS~J1843$-$033. The H.E.S.S. contours from 3$\sigma$ to 9$\sigma$ appear in cyan. (Right) Same as in the left panel; the SNR candidates found in the THOR radio survey \citep{Anderson:2017} are shown as green circles and the synchrotron-emitting SNR G28.6$-$0.1 and the energetic pulsar PSR~J1844$-$0346 are shown in white. The inset shows the 99 ks-long \chandra map of G28.6$-$0.1.}
    \label{fig:G28}
\end{figure*}

The unidentified source HESS~J1843$-$033 is defined as a large extended Gaussian component ($\sigma =0.24$\dg $\pm$ 0.06\dg) and results from the merging of two Gaussian components, which were previously detected by the HGPS pipeline, as illustrated in the HGPS significance map (Figure~\ref{fig:G28}, left). Several SNR candidates revealed in the THOR radio survey \citep{Anderson:2017} overlap with HESS~J1843$-$033 (Figure~\ref{fig:G28}, right). The radio SNR G28.6$-$0.1 and the energetic pulsar PSR~J1844$-$0346 \citep[$\dot{E}$ = 4.2 $\times$ 10$^{36}$ erg s$^{-1}$ and  $\tau_c$ = 11.6 kyr with unknown distance;][]{Clark:2017} are located within one of the previously detected Gaussian components. Figure~\ref{fig:G28} highlights the complexity of the region and shows a peak significance at nearly 6$\sigma$ at the position of the synchrotron-emitting SNR G28.6$-$0.1. 

The eastern and southern parts of the shell of G28.6$-$0.1 were first detected in radio, with a flux of $\approx$ 2 Jy at 1.4 GHz \citep{Helfand:1989}. \cite{Bamba:2001} revealed nonthermal emission from an extended \asca source spatially coincident with G28.6$-$0.1, with a power-law spectral index $\Gamma_{\rm{X}} = 2.1_{-0.4}^{+0.3}$, a column density of $N_{\rm{H}} = 2.6_{-0.6}^{+0.8} \times 10^{22}$ cm$^{-2}$ and a distance estimate between 6 kpc and 8 kpc. \chandra observations of G28.6$-$0.1 also revealed thermal and nonthermal X-ray emission components with a faint shell-like morphology \citep{Ueno:2003}. The association between G28.6$-$0.1 and the nearby X-ray pulsar PSR~J1844$-$0346 was considered owing to the column density of $N_{\rm{H}} = 5_{-3}^{+4} \times 10^{22}$ cm$^{-2}$, similar to that obtained for the SNR \citep{Zyuzin:2018}. The transverse velocity of the pulsar can be written as $v_{t} = 4.74 \mu D$ km s$^{-1}$, where $\mu$ is the proper velocity of the pulsar (mas/yr) and $D$ the distance (kpc) \citep{Lyne:1995}. Since $\mu$ has not been measured, we can approximate $\mu \approx \frac{\theta}{\tau_c}$, where $\theta$ is the angular distance between the current position of the pulsar and its place of birth and $\tau_c$ its characteristic age. Assuming a distance of 4.3 kpc \citep[estimated from the empirical relation obtained for $\gamma$-ray pulsars;][]{Saz:2010} and 6 kpc (the closest estimate for G28.6$-$0.1), the transverse velocity of the pulsar is $v_{t} \approx 1020$ km s$^{-1}$ and $\approx 1422$ km s$^{-1}$ \citep{Zyuzin:2018}, while typical pulsar velocities range between 400$-$500 km s$^{-1}$ \citep{Lyne:1995}. The association between PSR~J1844$-$0346 and G28.6$-$0.1 is therefore unlikely.

Using data from the TGSS, NVSS, and PMN surveys, we investigated the radio spectrum of G28.6$-$0.1 and we derived a spectral index of $\alpha = -0.44 \pm 0.01$, fully consistent with that usually measured for shell-type SNRs. The brightest parts of the shell are reported as sources in the TGSS catalog (at 150 MHz). By summing up their respective fluxes we obtained $\sim$ 8 Jy, which is broadly consistent with our estimate of 9.01 $\pm$ 0.17 Jy obtained for the entire SNR. At X-ray energies, the inset in Figure~\ref{fig:G28} (right) depicts the \chandra image, which shows a faint shell-like morphology, previously revealed by \cite{Ueno:2003}. No \fermi cataloged sources are reported close to the SNR. We also explored the archival multiwavelength data toward the pulsar PSR~J1844$-$0346 and we did not find any radio or X-ray counterpart that could indicate a possible PWN. Radio data from MAGPIS and infrared data from \ita{Spitzer} show a bright extended emission southeast of PSR~J1844$-$0346. The Simbad database indicates the presence of the star-forming region N49 containing \hii regions, whose distance is estimated to be 5.1 kpc or 9.7 kpc; the former is the most likely distance \citep[see][]{ Anderson_Bania:2009,Dirienzo:2012}. With an angular distance between N49 and PSR~J1844$-$0346 of $\theta$ = 181.5'', the proper velocity of the pulsar is $\mu = 15.65~t_{11.6}^{-1}$ mas yr$^{-1}$, assuming $t \approx \tau_c$ = 11.6 kyr. Taking the near distance to N49 ($d = 5.1$ kpc), the transverse velocity would be $v_t \sim 378~ d_{5.1}t_{11.6}^{-1}$ km s$^{-1}$, making the association of PSR~J1844$-$0346 with N49 more likely than with G28.6$-$0.1.

We shed light on a significant TeV excess that is spatially coincident with the synchrotron-emitting shell-type SNR G28.6$-$0.1. We investigated its radio spectrum and we derived a spectral index of $\alpha = -0.44 \pm 0.01$. Multiwavelength data do not show any emission from a potential PWN originating from the nearby and energetic pulsar PSR~J1844$-$0346. While the association between PSR~J1844$-$0346 and G28.6$-$0.1 was previously considered, we found that the pulsar more likely originates from the star-forming region N49. The radio and X-ray synchrotron-emitting SNR G28.6$-$0.1 is a promising source to be studied at HE and VHE since it should normally be able to accelerate particles up to $\gamma$-ray emitting energies in a leptonic scenario. More generally, the unidentified source HESS~J1843$-$033 is an interesting source to investigate at VHE, given the numerous overlapping SNR candidates. Among these, we note that there is another excess (at $\sim$4$-$5$\sigma$ in the HGPS maps) lying on the eastern part of the shell of G28.78$-$0.44. With unprecedented angular resolution and sensitivity, the next generation of Cherenkov telescopes CTA \citep{Science_CTA:2017} will give better insight into this region and might be able to confirm the VHE $\gamma$-ray emission from G28.6$-$0.1 (and potentially from G28.78$-$0.44 as well), making it a new TeV-emitting shell-type SNR.

\section{Conclusions}

Given the large number of HGPS sources that are not firmly identified, we developed a generic code aiming to constrain the origin of their TeV emission through the exploitation of multiwavelength data. The algorithm is based on an automatically retrieval of archival data from radio continuum, X-ray, infrared, and GeV instruments toward any region of the sky to search for faint counterparts through a careful visual inspection in every available image. The pipeline was completed by an association procedure using the catalogs of known objects (SNRs, PWNe, \hii regions, etc.) and those related to the instruments whose data are exploited. The main constraints on the origin of the TeV emission are obtained through the derivation of a radio spectral index that helps us disentangle thermal from nonthermal emission and through the estimate of the mean magnetic field under the assumption of a leptonic scenario. Finally, the \fermi cataloged source spectra are used to search for a smooth connection between GeV and TeV energies. This algorithm is well suited for isolated sources but faces some limitations for large and complex regions, although it gives an overall insight into the possible multiple components as the origin of the TeV emission.

We applied this pipeline on two unidentified sources reported in the HGPS: HESS~J1427$-$608 and HESS~J1458$-$608. We found extended nonthermal radio emission toward both sources and a mean magnetic field value consistent with that expected for evolved PWNe. \fermi data revealed a pulsar-like spectrum and a PWN-like spectrum in the vicinity of HESS~J1427$-$608 and HESS~J1458$-$608, respectively. We modeled the broadband nonthermal emission of both sources in a leptonic scenario with $B \lesssim 10$ $\mu$G, which is consistent with that obtained from ancient PWNe. We estimated the spin-down power and the characteristic age of the putative pulsars and we found that these are broadly in line with those of the population of known TeV PWNe. Deeper X-ray observations are necessary to potentially reveal a compact source in the vicinity of HESS~J1427$-$608. A detailed search for pulsation on 4FGL~J1427.8$-$6051, which exhibits a pulsar-like spectrum, could help reveal its nature. Follow-up X-ray observations could also reveal an extended source associated with HESS~J1458$-$608, and dedicated HE and VHE $\gamma$-ray data analyses are required to reveal in details the morphology of this unidentified VHE source.

We also shed light on a possibly significant, yet uncataloged, $\gamma$-ray excess in the HGPS data, located in the south of the unidentified source HESS~J1632$-$478, for which we found an extended radio counterpart with a complicated shell-like structure. The radio emission was found to be thermal and could be associated with a star-forming region, which could accelerate particles up to $\gamma$-ray emitting energies. We also highlighted another uncataloged $\gamma$-ray excess, that is spatially coincident with the synchrotron-emitting shell-type SNR G28.6$-$0.1, lying near the unidentified source HESS~J1833$-$043. Both sources have multiwavelength counterparts and could be revealed as new VHE sources in the future.

This code was developed with the purpose of applying a uniform method on the unidentified TeV sources to constrain the origin of their emission. Given the constraints obtained with archival data, this work also aims to advocate deeper multiwavelength observations, searches for pulsation, and dedicated analyses to be able to firmly identify TeV sources. \cite{Remy:2020} recently investigated blind search methods for VHE $\gamma$-ray source detection based on image processing and pattern recognition techniques and provided a list of promising HGPS TeV source candidates. Future work will include the application of this pipeline on these source candidates. If interesting multiwavelength counterparts are to be found, detailed VHE data re-analyses would then be warranted to firmly confirm these candidates as new VHE sources. Our next improvements of the code will include the use of \hi and CO data to assess the environment of the sources and the analysis of X-ray observations to derive better constraints on the synchrotron spectrum. The convolution of the radio and X-ray maps with the H.E.S.S. PSF would be also needed to quantify the spatial correlations at different wavelengths.

The next generation of Cherenkov telescopes CTA \citep{Science_CTA:2017} will certainly detect a significant number of new sources, leading to a higher degree of source confusion and hence, a larger number of unidentified sources than what is currently observed. \cite{Dubus:2013} estimated that 20$-$70 shell-type SNRs and 300$-$600 PWNe will be detected during the CTA Galactic Plane Survey. With significantly improved sensitivity and angular resolution compared to current Cherenkov instruments, CTA will provide new insights into the Galactic $\gamma$-ray sky and the synergy with multiwavelength observations will be necessary to constrain the origin of the observed TeV emission. The large field-of-view X-ray instrument eROSITA\footnote{\url{https://www.hs.uni-hamburg.de/hserosita/}}, recently launched, will soon give an unprecedented high-energy map up to 10 keV \citep{eROSITA:2012}, allowing us to search for faint X-ray counterparts along the whole Galactic plane. Data from the Square Kilometre Array \citep{SKA:2018} are also very promising to help identify $\gamma$-ray sources. Thus, future VHE observations and multiwavelength data exploitation supported by new generation instruments will probably reveal the nature of a significant number of unidentified TeV sources in order to assess their importance within the issue of the origin of the Galactic CRs.

\small{
\begin{acknowledgements}
We thank the anonymous referee for his/her comments that improved the manuscript. JD and MLG  acknowledge support from Agence Nationale de la Recherche (grant ANR-17-CE31-0014). GMRT is run by the National Centre for Radio Astrophysics of the Tata Institute of Fundamental Research. This research has made use of data from the Karl G. Jansky Very Large Array and from the MOST, which is operated by The University of Sydney with support from the Australian Research Council and the Science Foundation for Physics within The University of Sydney. The National Radio Astronomy Observatory is a facility of the National Science Foundation operated under cooperative agreement by Associated Universities, Inc. The research presented in this paper has used data from the Canadian Galactic Plane Survey, a Canadian project with international partners, supported by the Natural Sciences and Engineering Research Council. The Australia Telescope Compact Array and the Parkes radio telescope are part of the Australia Telescope National Facility which is funded by the Australian Government for operation as a National Facility managed by CSIRO. This research has made use of data from the Green Bank telescope, the Planck satellite, CHIPASS and the THOR survey. This research has made use of data obtained from the \chandra~Source Catalog, provided by the Chandra X-ray Center (CXC) as part of the Chandra Data Archive, and from the Suzaku satellite, a collaborative mission between the space agencies of Japan (JAXA) and the USA (NASA). This research has made use of data obtained from the 4XMM XMM-Newton serendipitous source catalogue compiled by the 10 institutes of the XMM-Newton Survey Science Centre selected by ESA. This work has made use of data supplied by the UK Swift Science Data Centre at the University of Leicester, and is based on observations with INTEGRAL, an ESA project with instruments and science data centre funded by ESA member states (especially the PI countries: Denmark, France, Germany, Italy, Switzerland, Spain), and with the participation of Russia and the USA. We have also used data from the NASA satellite NuSTAR and from ASCA, a collaborative mission between ISAS and NASA. This work has made use of the ROSAT Data Archive of the Max-Planck-Institut für extraterrestrische Physik (MPE) at Garching, Germany. This work is based in part on observations made with the Spitzer Space Telescope, which is operated by the Jet Propulsion Laboratory, California Institute of Technology under a contract with NASA. This research has made use of the SIMBAD database, operated at CDS, Strasbourg, France, and of data and/or software provided by the High Energy Astrophysics Science Archive Research Center (HEASARC), which is a service of the Astrophysics Science Division at NASA/GSFC.
\end{acknowledgements}}

\bibliographystyle{aa}
\bibliography{TeVsources}

\onecolumn
\begin{appendix}
\section{Instrument characteristics and catalogs}\label{sec:Appendix_A}

Table~\ref{tab:Radio_instru} gives the description of the radio instruments whose archival data are retrieved through the pipeline.

\begin{table*}[h!]
\centering
\tiny{
\begin{tabular}{lccccccc}
\hline
\hline
Instruments     & Type & Coverage & Frequency (GHz) & Sensitivity (mJy) & PSF (FWHM) &    Data links & References \\ 
\hline
TGSS                        &  I        &       $\delta > -53^{\circ}$ &        0.150  & 24.5 & $\sim 0.45'$* & [1]* &  \cite{TGSS_Intema:2017} \\ 
MGPS-2                          & I             & $245^{\circ}<l<365^{\circ}$         & 0.843 & 8     & 45'' & [2]* & \cite{MGPS2_Murphy:2007} \\ 
                                &               &       $b < 10^{\circ}$        &       &       &               &       \\
THOR                            &       I       & $14.5^{\circ} \leq l \leq 67.4^{\circ}$   &       1.06/1.31/1.44 &        $\sim$ 6        &       0.42' & [3] &  \cite{THOR_Beuther:2016} \\
                                &               &       $|b| < 1.25^{\circ}$                                            & 1.69/1.82  &  $\sim$ 2.5 &   & \\
CHIPASS                 & S     &       $\delta < 25^{\circ}$ & 1.4     &       90 &       14.4' & [4] & \cite{CHIPASS_Calabretta:2014} \\                                 
VGPS                                    & I     & $18^{\circ}<l<67^{\circ}$             & 1.4 & 20 & 1'  & [5] & \cite{VGPS_Stil:2006} \\ 
                        &  & $|b| < 1.3-2.3^{\circ}$    &       &       &       &       \\
NVSS                                    & I             &  $\delta \geq - 40^{\circ}$     &       1.4  &  2.5 & 45'' & [6]* &     \cite{NVSS_Condon:1998} \\                              
CGPS                                    & I     &       $74.2^{\circ}<l<147.3^{\circ}$  &       1.42                            & 1.3 &   $\sim$ 1'*      & [7]* & \cite{CGPS_Taylor:2003} \\
                                        &               &       $-3.6^{\circ} < b < 5.6^{\circ}$      &       &       &       &       \\
SGPS                                    & I             &       $253^{\circ} \leq l \leq 358^{\circ}$         &      1.42 & 1  & 1.67' & [8] & \cite{SGPS_McClure:2005} \\      
                                        &               &       $|b| < 1.5^{\circ}$     &       &       &       &       \\
Parkes                                  & S     & $238^{\circ} \leq l \leq 365^{\circ}$            & 2.4   &60     &       10.62' & [9]* & \cite{Parkes_Duncan:1995} \\      
                                        &               &       $|b| \leq 5^{\circ}$      &       &       &       &       \\
PMN                                             &       S       & $-87.5^{\circ} < \delta < 10^{\circ}$  & 4.85  &       20 & 4.9' & [10]* & \cite{PMN_Griffith:1993} \\      
GBT/GPA                         & S     &  $-15^{\circ}<l<255^{\circ}$          &       8.35  &       900  & 11.17' & [11]* & \cite{GBT_GPA_Langston:2000} \\                
                                &  &    $|b| < 5^{\circ}$                                       & 14.35 & 2500 & 8.0'             &\\
87GB                                    & S & $0^{\circ} < \delta < 75^{\circ}$ & 4.85  & 18 & $\sim$ 3.72' & [12]*&      \cite{87GB_Gregory:1991} \\     
MAGPIS                          & I     & $3.6^{\circ}<l<33.2^{\circ}$  &       0.325  &      1.5  &  1'  & [13]* & \cite{MAGPIS_Helfand:2006} \\             
                                &       &       $|b| < 2^{\circ}$       &       &       &               &\\
                            & I         & $5^{\circ}<l<48.5^{\circ}$    &       1.4 &        1.5  & $\sim 0.12'$    &\\     
                            &   &       $|b| < 0.8^{\circ}$     &       &       &       &\\
                                & I     & $350^{\circ}<l<42^{\circ}$    &       5  &      1.5  &   0.03' &\\                      
                                &       &       $|b| < 0.4^{\circ}$     &       &       &       &       \\
\hline
\end{tabular}
}
\caption{Properties of the radio surveys. The type "S" and "I" denotes a "single-dish telescope" or "interferometers", respectively. The point-source sensitivity is given, as well as the FWHM. If the FWHM is asymmetric, the major axis is given. For the TGSS and the CGPS, the FWHM depends on the declination (indicated by the asterisk). The links to retrieve these data are labeled with numbers and given below. Numbers followed by an asterisk indicate that a source catalog, associated with the instrument, is available and used.}
\label{tab:Radio_instru}
\end{table*}

Archival data are retrieved at the following links: 
\begin{itemize}
    \item TGSS: [1] \url{http://vo.astron.nl/tgssadr/q_fits/imgs/form} \\
    \item MGPS-2: [2] \url{http://www.astrop.physics.usyd.edu.au/mosaics/Galactic} \\
    \item THOR: [3] \url{http://www2.mpia-hd.mpg.de/thor/DATA/www/} \\
    \item CHIPASS: [4] \url{http://www.atnf.csiro.au/research/CHIPASS/} \\
    \item VGPS: [5] \url{http://www.ras.ucalgary.ca/VGPS/VGPS_data.html} \\
    \item NVSS: [6] \url{http://www.cv.nrao.edu/nvss/} \\
    \item CGPS: [7] \url{http://www.cadc-ccda.hia-iha.nrc-cnrc.gc.ca/AdvancedSearch} \\
    \item SGPS: [8] \url{http://www.atnf.csiro.au/research/HI/sgps/fits_files.html} \\
    \item Parkes: [9] \url{http://www.atnf.csiro.au/research/surveys/2.4Gh_Southern/data.html} \\
    \item PMN: [10] \url{ftp://ftp.atnf.csiro.au/pub/data/pmn/maps/PMN/} \\
    \item GBT/GPA: [11] \url{http://www.gb.nrao.edu/~glangsto/gpa/} \\
    \item 87GB: [12] \url{ftp://ftp.atnf.csiro.au/pub/data/pmn/maps/87GB/} \\
    \item MAGPIS: [13] \url{https://third.ucllnl.org/gps/index.html} \\
\end{itemize}

\newpage
Table~\ref{tab:Xray_instru} summarizes the description of the X-ray instruments whose archival data are retrieved through the pipeline. We automatically extract all the observations whose center is comprised within an angular distance from the source of interest. For ROSAT/PSPC and \xmm, we create background-subtracted and exposure-corrected maps, while the count maps of \swift/XRT, ASCA and \suzaku are only corrected from the exposure because the corresponding background maps are not available. Except for large field-of-view instruments (such as ROSAT/PSPC, {\it Integral}/IBIS-ISGRI, and \swift/BAT), we create a mosaic of these images in case of multiple observations. For \chandra data, we directly retrieve stacked and subtracted-background images (combining multiple observations), which are also corrected from the exposure and the vignetting. For {\it Integral}/IBIS-ISGRI and \swift/BAT, we extract a subregion of the flux, error flux, and significance maps.

\begin{table*}[h!]
\centering
\begin{minipage}{20cm}
\tiny{
\begin{tabular}{lcccccc}
        \hline
        \hline
                Satellite               &                       Detector                         &                       Field of view                   &         Energy  (keV)                                   &               PSF (FWHM)          & Data link &  Catalog names \\ 
                \hline
                \chandra                &                       ACIS                                            &                                 $17'$ $\times$ $17'$    &       0.5 -- 7.0                                         &                       $1''$           & [1]*    & CSC2\footnote{\url{http://cxc.cfa.harvard.edu/csc2/}}  \\             
                \suzaku                 &                       XIS                                             &                                 $19'$ $\times$ $19'$    &       0.2 -- 12.0                                     &                         $< 1.5'$                & [2] & --      \\
                \ita{XMM-Newton}         &              EPIC (MOS1-2)           &                                 $30'$ $\times$ $30'$            &       0.2 --      12.0                                    &                       $6''$    & [3]*  & 4XMM-DR9\footnote{\url{http://xmmssc.irap.omp.eu/Catalogue/4XMM-DR9/4XMM_DR9.html}, \cite{Webb:2020}} \\                                    
                                                         &              EPIC (PN)                                    &                               $30'$ $\times$ $30'$          &       0.2 --  12.0                                    &                         $6''$                           \\              
                \asca                   &                       SIS                                             &                                 $22'$ $\times$ $22'$            &       0.4 -- 12.0                                 &       $1'$                                                     & [4]*  & GIS/SIS/Galactic Plane Survey\footnote{\url{https://heasarc.gsfc.nasa.gov/W3Browse/all/ascasis.html} (same for ascagis.html and ascagps.html)}\\                     
                                                        &               GIS                                             &              $20'$ $\times$ $20'$               &  0.6 -- 12.0                                  & $1'$                                                    \\                      
                \swift                  &                       XRT                                             &                         $23.6'$ $\times$ $23.6'$        &       0.2 -- 10.0                                         &                       $18''$*         & [5]*  & 1SWXRT/2SXPS\footnote{\url{http://www.asdc.asi.it/1swxrt/}, \url{http://www.swift.ac.uk/2SXPS/}, \cite{1SWXRT_2013}, \cite{2SXPS_2020} } \\
                                                        &                       BAT                                             &                               $5400'$ $\times$ $3000'$                &       14 -- 195                                       &                         $17'$   & [6]*  & BS105 months\footnote{\url{https://swift.gsfc.nasa.gov/results/bs105mon/}}\\  
                Integral                        &                       IBIS                                            &                 $1140'$ $\times$ $1140'$                &       17 -- 60                                                &                         $12'$    & [7]* & 8-1 yr All sky and 9-14 Galactic sky\footnote{\url{https://heasarc.gsfc.nasa.gov/W3Browse/integral/ibiscat.html},\url{https://heasarc.gsfc.nasa.gov/w3browse/all/intibisvhd.html},\url{https://heasarc.gsfc.nasa.gov/w3browse/all/intibisgal.html},\url{http://hea.iki.rssi.ru/integral/fourteen-years-galactic-survey/}}        \\              
                NuSTAR          &                       --                                                      &                         $13'$ $\times$ $13'$            &       5 -- 80                                                 &                         $7.5''$         & [8]* & Survey source catalog \citep{Nustar_cat:2017}          \\      
                ROSAT                   &                       PSPC                                    &                         $114'$ $\times$ $114'$  &               0.1 -- 2.4                              &                         $45''$                  & [9]* & 1RXS \citep{1RXS:1999} \\
                & & & & & & 2RXS \citep{2RXS:2016}                                              \\              
\hline
\end{tabular}
}\end{minipage}{20cm}
\caption{Properties of the X-ray instruments. The \swift/XRT PSF (noted with an asterisk) corresponds to the half-power diameter. The links to retrieve these data are labeled with numbers and given below. Numbers followed by an asterisk indicate that a source catalog, associated with the instrument, is available and used.}
\label{tab:Xray_instru}
\end{table*}

Archival X-ray data are retrieved through the following links:

\begin{itemize}
    \item \ita{Chandra}: [1] \url{http://cxc.cfa.harvard.edu/csc2/data_products/} \\
    \item \suzaku: [2]  \url{http://www.darts.isas.jaxa.jp/pub/suzaku/} \\
    \item \xmm: [3]  \url{http://nxsa.esac.esa.int/nxsa-sl/} \\
     \item \asca: [4] \url{https://darts.isas.jaxa.jp/pub/asca/data/} \\
    \item \swift/XRT: [5] \url{ftp://legacy.gsfc.nasa.gov/swift/data/} \\
    \item \swift/BAT: [6] \url{https://skyview.gsfc.nasa.gov/} \\
    \item Integral: [7] \url{http://hea.iki.rssi.ru/integral/fourteen-years-galactic-survey/} \\
    \item NuSTAR: [8] \url{https://heasarc.gsfc.nasa.gov/FTP/nustar/data/} \\
    \item ROSAT: [9] \url{ftp://ftp.mpe.mpg.de/rosat/archive/} \\
\end{itemize}

To complete our data set, we extract infrared data from Spitzer/GLIMPSE at 3.6, 4.5, 5.8, 8, 21, 24, 870, and 1100 $\mu$m (available on the MAGPIS website \url{https://third.ucllnl.org/gps/}). Using ten years of \fermi data, we also created a binned count map with energies between 10 and 500 GeV, setting a pixel size of 0.05\dg and 10 energy bins (version 1.0.10 of the \courrier{Fermitools}). The map encompasses the HGPS with latitude $|b| \leq 5$\dg. The \courrier{SOURCE} event class was selected, which is a compromise between background rejection and statistics and we imposed a maximum zenith angle on the photon arrival direction of 90\dg to reduce the contamination of the Earth limb. We excluded time intervals during which the satellite passed through the South Atlantic Anomaly and when the rocking angle was more than 52\dg. We modeled the contribution of the Galactic and isotropic diffuse emissions using the files \courrier{gll\_iem\_v07.fits} and \courrier{iso\_P8R3\_SOURCE\_V2.txt}\footnote{Available at \url{https://fermi.gsfc.nasa.gov/ssc/data/access/lat/BackgroundModels.html}} (with the normalizations fixed to 1 and the spectral index to 0) using the instrument response functions \courrier{P8R3\_SOURCE\_V2}, and we created a residual count map between 10 GeV and 500 GeV.

\newpage
Table~\ref{tab:catalogs} reports the catalogs of known objects used for the association procedure.

\begin{table*}[h!]
\centering
\begin{minipage}{20cm}
\begin{tabular}{lcc}
        \hline
        \hline
                Objects                                 &       Name and details       &         References                      \\
                \hline
                Pulsars                         &       ATNF (version 1.58)       & \cite{ATNF:2005}\footnote{\url{http://www.atnf.csiro.au/people/pulsar/psrcat/}} \\
                Galactic SNRs               &   Green catalog   &               \cite{Green:2017}                               \\
                SNRs/PWNe                   &   SNR cat         &      \cite{Ferrand:SNRcat}\footnote{\url{http://www.physics.umanitoba.ca/snr/SNRcat}}                     \\
                Galactic SNR candidates     & found with THOR data  & \cite{Anderson:2017} \\
                \fermi sources              &   4FGL (50 MeV$-$1 TeV)  &    \cite{4FGL}           \\
                                            &   3FGL (0.1$-$300 GeV)  &          \cite{3FGL}         \\
                                            &  2FHL (50 GeV$-$2 TeV)    &  \cite{2FHL}     \\
                                            &  3FHL (10 GeV$-$2 TeV) &   \cite{3FHL:2017}    \\
                 \fermi extended sources    & FGES ($>10$ GeV)         &  \cite{FGES:2017} \\
                 \fermi pulsars                &   2PC                 &      \cite{2PC:2013}         \\
                HAWC sources                &  2HWC                 &    \cite{2HWC:2017}                       \\
                \hii regions                & obtained with WISE data   &  \cite{Anderson_WISE:2014}\footnote{\url{http://astro.phys.wvu.edu/wise/}}         \\
                Molecular clouds            &  Galactic MCs                     &   \cite{Rice:2016}                \\
                Galactic O-Stars            & GOSC                      &    \cite{GOSC:2013}               \\
\hline
\end{tabular}
\end{minipage}{20cm}
\caption{Catalogs used for the association procedure.}\label{tab:catalogs}
\end{table*}

\section{Radio spectral index derivation}\label{sec:Appendix_B}

To estimate the radio spectral index, we calculate the flux at different frequencies with the available archival continuum data. We mask all the cataloged radio sources in the images that are located outside the flux extraction region (called $R_{\rm{ON}}$), which is defined after a visual inspection of the images. The maps are in Jy/beam (hereafter $\Phi$) and the associated beam is the full width at half maximum (FWHM) for which the effective area is given by\begin{equation}
B_{\rm{area}} = \frac{2 \pi ab}{(2 \sqrt{2 \ln{2}})^2} \hspace{0.3cm} \text{deg}^2
,\end{equation}
where $a$ and $b$ are the half major and minor axes of the instrument PSF. The number of pixels per beam is thus written as
\begin{equation}
N_{\rm{pix, psf}} = \frac{B_{\rm{area}}}{c_1 c_2}
,\end{equation}
where $c_1$ and $c_2$ are the pixel size in degrees in both directions. The mean background $B_{0}$ and the standard deviation (rms) are obtained by a Gaussian fit on the flux distribution of unmasked pixels outside the ON region. The total flux in the ON region and its associated error are given by
\begin{equation}
F_{\rm{tot}} = \frac{\sum_{i}^{N_{\rm{pix, on}}} \Phi_i}{N_{\rm{pix, psf}}} \hspace{0.3cm} \text{Jy} 
,\end{equation}
\begin{equation}
\sigma_{\rm{tot}} = \text{rms} \times \sqrt{\frac{N_{\rm{pix, on}}}{N_{\rm{pix, psf}}}} \hspace{0.3cm} \text{Jy} \\
,\end{equation}
where $N_{\rm{pix, on}}$ is the total number of pixels inside the ON region and $\Phi_i$ the flux in units of mJy/beam per pixel. The total background in the ON region is written as
\begin{equation}
B_{\rm{tot}} = B_{0} \times \frac{N_{\rm{pix, on}}}{N_{\rm{pix, psf}}} \hspace{0.3cm} \text{Jy}. \\
\end{equation}
The source flux in the ON region is thus $F_{\rm{ON}} = F_{\rm{tot}} - B_{\rm{tot}}$. If the significance is below 3$\sigma$, we derive an upper limit at the 3$\sigma$ confidence level as $\rm{UL} = 3$ $\times$ $\sigma_{\rm{tot}}$. We then fit the calculated fluxes at different frequencies with a power law. We only consider the data points (not the upper limits) for the fit, since in some cases the source masking in radio maps with moderate PSF can lead to an underestimation of upper limits. 

The radio flux is expressed as $S_{\nu} \propto \nu^{\alpha}$, where $\nu$ and $\alpha$ are the frequency and  spectral index, respectively. The spectral index $\alpha$ allows us to disentangle thermal from nonthermal emission for which we have $\alpha \gtrsim 0$ and $\alpha \lesssim 0,$ respectively. Radio spectral index of SNRs has a mean value of $\alpha \sim -0.5/-0.4$ \citep{Green:2017}, while for PWNe we have $\alpha \sim -0.3$/$0$ \citep{deJager:2009}.

\begin{figure*}[th!]
\centering
\includegraphics[scale=0.4]{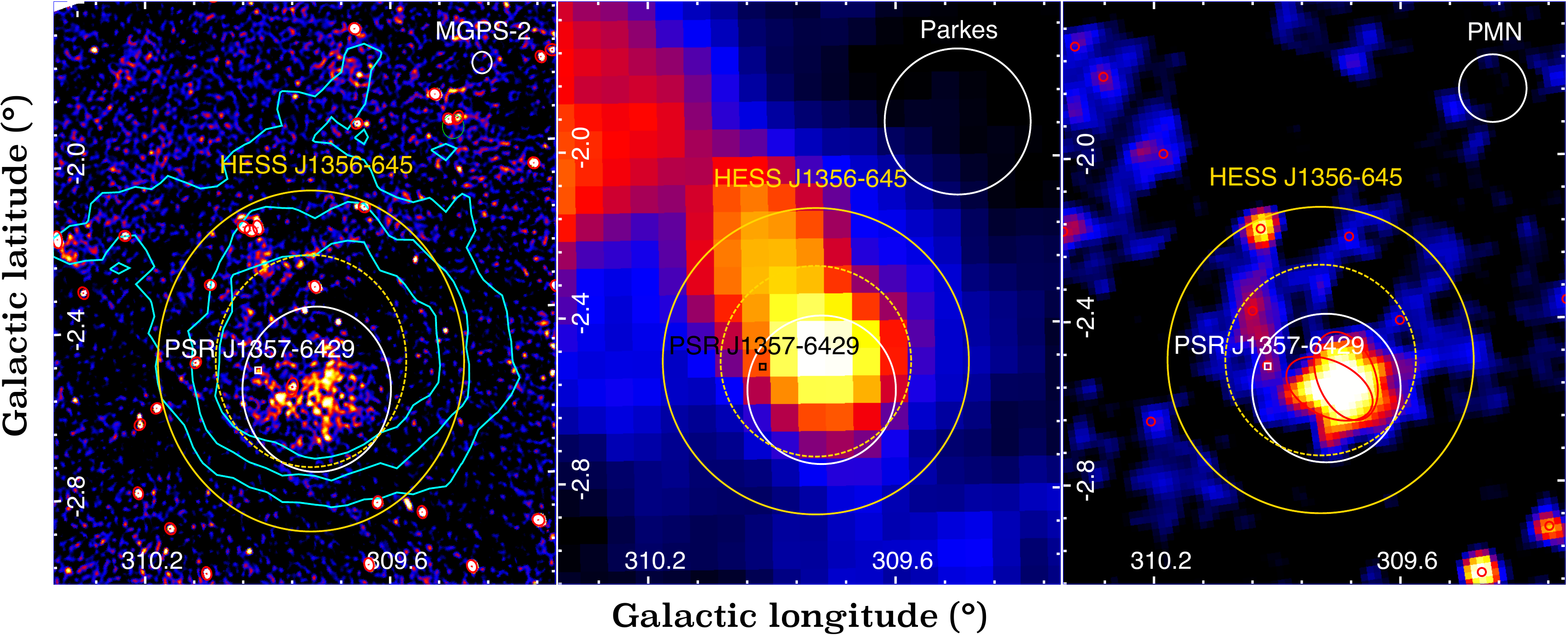}
\caption{MGPS-2, Parkes, and PMN maps of HESS~J1356$-$645. The ON region is represented by a white circle and the pulsar position is depicted by a white/black square. The PSF (FWHM) of each instrument is illustrated at the top right of the images. The cataloged radio sources are represented by red circles. The dashed and solid yellow circles indicate the Gaussian $\sigma$ extent and the flux extraction radius of HESS~J1356$-$645, respectively, as defined in the HGPS. The blue contours in the MGPS-2 map correspond to 3$\sigma$, 5$\sigma$, and 7$\sigma$ in the HGPS significance map.}
\label{fig:map_1356}
\vspace{1cm}
\begin{tabular}{l|ccccc}
        \hline
        \hline
&       MGPS-2          &               CHIPASS                                         &               Parkes                                         &                               PMN                                                  \\
        \hline
        Frequency (GHz)                         &               0.843           &                         1.4                                                             &               2.4                                                 &                               4.85                                                         \\
        rms (mJy/beam)                                  &               1.76                 &                       637.2                                                           &                       235.1                                           &                               10.57                                           \\
        Flux density (mJy)                      &        538.6 $\pm$ 42.3               &           $< 2822.0$                           & $< 1 296.6$    &                              509.4 $\pm$ 39.7                              \\
      \hline
\end{tabular}
\caption{Estimated fluxes with the associated statistical errors. The upper limits are given at the 3$\sigma$ confidence level.}
\label{tab:fit_radio_1356}
\vspace{1cm}
\includegraphics[scale=0.43]{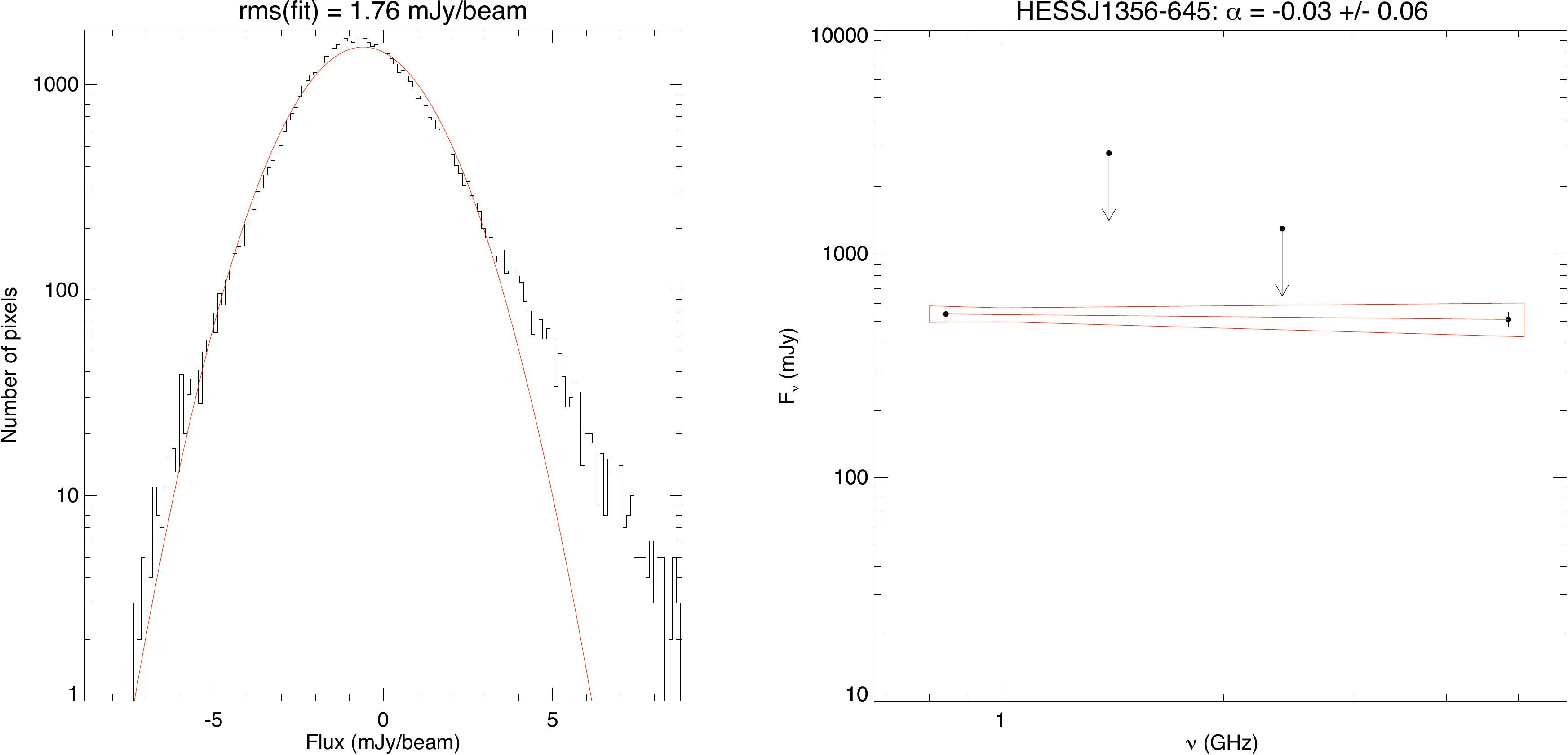}
\caption{(Left) Distribution of the flux in the MGPS-2 unmasked pixels located outside of the ON region and used for the background estimation. The best-fit Gaussian is represented in red. (Right) Calculated fluxes (black points) with the best-fit power law overlaid in red. The upper limits (black arrows) are given at the 3$\sigma$ confidence level and are not considered in the fit.}
\label{fig:ind_spec_radio_1356}
\end{figure*}

This method was used to identify the radio counterpart of the PWN HESS~J1356$-$645 in \cite{HESSJ1356_2011}, for which the derived spectral index was $\alpha = -0.01 \pm 0.07$. Below we illustrate the method, which is now implemented in a generic way involving more radio instruments, on the PWN HESS~J1356$-$645. The available data concerns those of the MGPS-2, Parkes and PMN instruments. Figure~\ref{fig:map_1356} depicts the radio maps and Table~\ref{tab:fit_radio_1356} reports the fluxes and upper limits obtained within the ON region defined by the white circle in Figure~\ref{fig:map_1356}. As shown, the pixel masking in the CHIPASS and Parkes data (with beams as large as $14.4'$ and $10.6'$, respectively) prevents us from deriving a meaningful flux due to a too small number of remaining unmasked pixels. The source masking is then the main limitation of this method\footnote{The source masking sometimes prevented us from using data (for the flux calculation) from relatively low angular resolution instruments such as CHIPASS and Parkes toward HESS~J1427$-$608, HESS~J1458$-$608, and HESS~J1632$-$478.}. Figure~\ref{fig:ind_spec_radio_1356} depicts the distribution of the unmasked pixels in the OFF region (used for the background estimation) of the MGPS-2 data (left), and the SED with the best-fit power law on the calculated fluxes. We found a radio spectral index of $\alpha = -0.03 \pm 0.06$ that is compatible with the value obtained in \cite{HESSJ1356_2011}. The difference can be explained by a slightly different background estimation method and a small difference in the size and position adopted for the ON region.

\newpage
\section{Mean magnetic field estimation}\label{sec:Appendix_C}

We use ROSAT/PSPC data in the 0.9--2.4 keV band to estimate the flux (or the flux upper limit) and to constrain the value of the mean magnetic field, given the measured flux at TeV energies. We take advantage of this large field-of-view all-sky instrument, which covers the entire HGPS region. We use the 1RXS catalog reporting the brightest sources detected with ROSAT/PSPC \citep{1RXS:1999} and the 2RXS catalog \citep{2RXS:2016} to mask the sources located outside the ON region. We use the \ita{ring background} and the \ita{reflected background} methods, and we calculate the significance following the prescription in \cite{Li_Ma:1983}, using a correlation radius equal to the flux extraction radius defined in the HGPS (\rspec). For the \ita{ring background} method, the OFF region is defined between $R_{\rm{in}}$ = \rspec + 1 pixel and $R_{\rm{out}}$ = $\sqrt{(10R_{\rm{spec}}^2 + R_{\rm{in}}^2)}$\dg so that the OFF to ON area ratio amounts to $\sim$ 10 for large sources. For the \ita{reflected background} method, we use four circular regions (with a radius \rspec) located at an angular distance of 3 $\times$ \rspec from the source center. If the X-ray emission is not significant, an upper limit on the flux is calculated at a 5$\sigma$ confidence level. The X-ray flux is simulated using the HEASARC tools \courrier{WebPimms}\footnote{\url{https://heasarc.gsfc.nasa.gov/Tools/w3pimms_help.html}} and defined as
\begin{equation}
F_{\rm{obs}} = F_0 \times e^{-\sigma(E)N_{\rm{H}}} \hspace{1cm} \text{with} \hspace{1cm} F_0 = N_0 \bigg ( \frac{E}{E_0} \bigg )^{-\Gamma}
,\end{equation}
where $F_0$ is the non-absorbed flux, $\sigma(E)$ the photoelectric absorption cross section (in units of cm$^2$) and $N_{\rm{H}}$ the column density (in units of cm$^{-2}$). For each simulated flux, a combination of ($\Gamma$, $N_{\rm{H}}$) corresponds to a count rate between 0.9 and 2.4 keV that we simulate in the image with Monte-Carlo sample. The simulation stops when the significance seen in the ROSAT/PSPC data is reached (or when the 5$\sigma$-threshold is reached in case of a non-detection) and gives the X-ray flux (or upper limit on flux) in the parameter space ($\Gamma$, $N_{\rm{H}}$).

Then, we estimate the mean magnetic field assuming a one-zone model and that electrons radiating synchrotron and IC emission (scattering on the cosmic microwave background) are in the Thompson regime ($\Gamma_{\rm{sync}} = \Gamma_{\rm{IC}}=\Gamma$, where $\Gamma$ is the photon spectral index). If we describe the electron spectrum with a power law ($K \gamma^{-p}$, $\gamma$ and $p$ being the Lorentz factor and the electron spectral index), the ratio of the synchrotron to IC fluxes is
given by\begin{equation}
\frac{F_{\rm{sync}}}{F_{\rm{IC}}} = \frac{U_{\rm{B}}}{U_{\rm{CMB}}} \times \bigg ( \frac{\gamma_{\rm{2,sync}}^{3-p} - \gamma_{\rm{1,sync}}^{3-p}}{\gamma_{\rm{2,IC}}^{3-p} - \gamma_{\rm{1,IC}}^{3-p}} \bigg )
\label{eq:flux_ratio}
\end{equation}
where $U_{\rm{B}} \sim 2.5 B_{-5}^2$ eV cm$^{-3}$ and $U_{\rm{CMB}}$ = 0.26 eV cm$^{-3}$ are the magnetic field and cosmic microwave background energy densities respectively (with $B_{-5} = \frac{B}{10 \hspace{0.1cm} \mu \rm{G}}$). The synchrotron and IC radiations occur at the characteristic energies of
\begin{equation}
\hspace{2cm}
\gamma_{\rm{sync}} \simeq 1.4 \times 10^8 E_{\rm{sync,keV}}^{1/2} B_{-5}^{-1/2} \hspace{1cm} \text{and}  \hspace{1cm} \gamma_{\rm{IC}} \simeq 3.6 \times 10^7 E_{\rm{IC,TeV}}^{1/2}  
\end{equation}
Using the above equations, the mean magnetic field can be expressed as
\begin{equation}
B_{-5} \backsimeq \text{G}(\Gamma) \times \bigg (\frac{F_{\rm{sync}}}{F_{\rm{IC}}} \times \frac{E_{\rm{2,IC, TeV}}^{2-\Gamma} - E_{\rm{1,IC, TeV}}^{2-\Gamma}}{E_{\rm{2,sync, keV}}^{2-\Gamma} - E_{\rm{1,sync, keV}}^{2-\Gamma}} \bigg )^{1/ \Gamma}
\label{eq:B_ROSAT}
\end{equation}
where $F$ is the flux between $E_{\rm{1}}$ and $E_{\rm{2}}$, $\text{G}(\Gamma) \simeq (0.1 \times 15^{\Gamma-2})^{1/ \Gamma}$ and $\Gamma = (p + 1) / 2$. If $\Gamma=2$, the mean magnetic field is simply expressed as
\begin{equation}
\frac{F_{\rm{sync}}}{F_{\rm{IC}}} \simeq 10 B_{-5}^2
\end{equation}

We can thus estimate the mean magnetic field using the X-ray flux derived with ROSAT/PSPC data and the TeV flux and spectral index reported in the HGPS. Below the method is applied on the SNR RX~J1713.7$-$3946 for illustration.

\begin{figure*}
    \centering
    \includegraphics[scale=0.48]{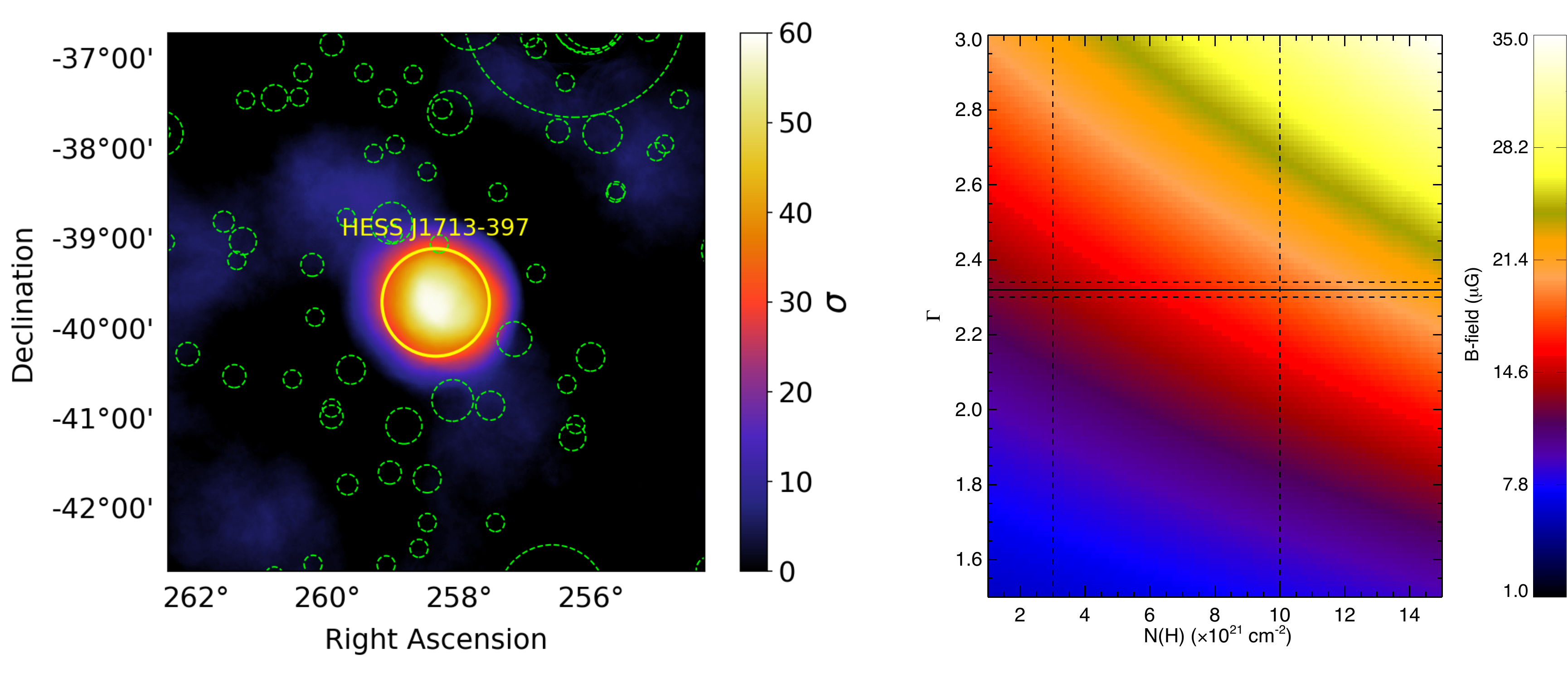}
    \caption{(Left) ROSAT/PSPC significance map using the \ita{reflected background} method and a correlation radius of \rspec = 0.6\dg (represented by a yellow circle). The dashed green circles are the 1RXS and 2RXS sources that have been masked in the ROSAT/PSPC count map (0.9 $-$ 2.4 keV). (Right) Mean magnetic field as a function of the TeV photon spectral index and the column density. The full and dashed horizontal lines are the TeV spectral index and its associated statistical errors, as reported in the HGPS. The vertical dashed lines correspond to the column density range derived from \cite{Sano:2015}.}
    \label{fig:Bfield}
\end{figure*}

The spectral extraction radius of the SNR RX~J1713.7$-$3946 is \rspec = 0.6\dg and the spectral parameters reported in the HGPS are $F_{1\hspace{0.1cm}\rm{TeV}} = (2.13 \pm 0.94) \times 10^{-11}$ cm$^{-2}$ s$^{-1}$ TeV$^{-1}$ and $\Gamma_{\rm{TeV}} = 2.32 \pm 0.02$. We masked in the ROSAT/PSPC images all the 1RXS and 2RXS cataloged sources whose centroid are located outside \rspec. Figure~\ref{fig:Bfield} (left) shows the significance map using the \ita{reflected background} method, which is similar to that obtained using the \ita{ring background} method. As expected, the significance is well above the 5$\sigma$ threshold. The mean value of the magnetic field is given in Figure~\ref{fig:Bfield} (right) with respect to the TeV photon spectral index and the column density. Following \cite{Sano:2015}, who estimated the variation of the column density over the entire SNR, we took $N_{\rm{H}} = (0.3-1.0) \times 10^{22}$ cm$^{-2}$. Within these TeV spectral index and column density ranges, we determined a mean magnetic field of $B = [10.48-16.37]$ $\mu$G, which is consistent with $B = 14.26 \pm 0.16$ $\mu$G, obtained when modeling the broadband nonthermal emission of the SNR with a leptonic model \citep{RXJ:2018}.

\end{appendix}

\end{document}